\documentclass[sigconf]{acmart}

\usepackage{caption}
\usepackage{subcaption}
\usepackage{epstopdf}
\usepackage{enumitem}
\usepackage{algorithm}
\usepackage{algorithmic}
\usepackage{multirow}

\newtheorem{definition}{Definition}

\setcopyright{rightsretained}
\acmDOI{10.1145/3269206.3271739}
\acmISBN{978-1-4503-6014-2/18/10}
\acmConference[CIKM '18]{The 27th ACM International Conference on Information and Knowledge Management}{October 22--26, 2018}{Torino, Italy}
\acmBooktitle{The 27th ACM International Conference on Information and Knowledge Management (CIKM '18), October 22--26, 2018, Torino, Italy}
\acmYear{2018}
\copyrightyear{2018}
\acmPrice{15.00}

\begin{document}
\title[RippleNet for Recommender Systems]{RippleNet: Propagating User Preferences on the Knowledge Graph for Recommender Systems}

\author[H. Wang et al.]{Hongwei Wang$^{1,2}$, Fuzheng Zhang$^3$, Jialin Wang$^{4}$, Miao Zhao$^{4}$, Wenjie Li$^4$, Xing Xie$^2$, Minyi Guo$^1$}
\authornote{M. Guo is the corresponding author. This work was partially sponsored by the National Basic Research 973 Program of China under Grant 2015CB352403.}
\affiliation{$^1$Shanghai Jiao Tong University, wanghongwei55@gmail.com, guo-my@cs.sjtu.edu.cn}
\affiliation{$^2$Microsoft Research Asia, xingx@microsoft.com, $^3$Meituan AI Lab, zhangfuzheng@meituan.com}
\affiliation{$^4$The Hong Kong Polytechnic University, \{csjlwang, csmiaozhao, cswjli\}@comp.polyu.edu.hk}

\begin{abstract}
	To address the sparsity and cold start problem of collaborative filtering, researchers usually make use of side information, such as social networks or item attributes, to improve recommendation performance.
	This paper considers the knowledge graph as the source of side information.
 	To address the limitations of existing embedding-based and path-based methods for knowledge-graph-aware recommendation, we propose \textit{RippleNet}, an end-to-end framework that naturally incorporates the knowledge graph into recommender systems.
 	Similar to actual ripples propagating on the water, RippleNet stimulates the propagation of user preferences over the set of knowledge entities by automatically and iteratively extending a user's potential interests along links in the knowledge graph.
 	The multiple "ripples" activated by a user's historically clicked items are thus superposed to form the preference distribution of the user with respect to a candidate item, which could be used for predicting the final clicking probability.
 	Through extensive experiments on real-world datasets, we demonstrate that RippleNet achieves substantial gains in a variety of scenarios, including movie, book and news recommendation, over several state-of-the-art baselines.
\end{abstract}


\maketitle

\subsection*{\small{ACM Reference Format:}}
\vspace{-0.05in}
	{\small
		Hongwei Wang, Fuzheng Zhang, Jialin Wang, Miao Zhao, Wenjie Li, Xing Xie, and Minyi Guo.
		2018.
		RippleNet: Propagating User Preferences on the Knowledge Graph for Recommender Systems.
		In \textit{The 27th ACM International Conference on Information and Knowledge Management (CIKM '18), October 22--26, 2018, Torino, Italy}.
		ACM, New York, NY, USA, 10 pages.
		https://doi.\\org/10.1145/3269206.3271739
	}

\section{Introduction}
	The explosive growth of online content and services has provided overwhelming choices for users, such as news, movies, music, restaurants, and books.
	Recommender systems (RS) intend to address the information explosion by finding a small set of items for users to meet their personalized interests.
	Among recommendation strategies, \textit{collaborative filtering} (CF), which considers users' historical interactions and makes recommendations based on their potential common preferences, has achieved great success \cite{koren2009matrix}.
	However, CF-based methods usually suffer from the sparsity of user-item interactions and the cold start problem.
	To address these limitations, researchers have proposed incorporating \textit{side information} into CF, such as social networks \cite{jamali2010matrix}, user/item attributes \cite{wang2018shine}, images \cite{zhang2016collaborative} and contexts \cite{sun2017collaborative}.
	
	Among various types of side information, \textit{knowledge graph} (KG) usually contains much more fruitful facts and connections about items.
	A KG is a type of directed heterogeneous graph in which nodes correspond to \textit{entities} and edges correspond to \textit{relations}.
	Recently, researchers have proposed several academic KGs, such as NELL\footnote{\url{http://rtw.ml.cmu.edu/rtw/}}, DBpedia\footnote{\url{http://wiki.dbpedia.org/}}, and commercial KGs, such as Google Knowledge Graph\footnote{\url{https://www.google.com/intl/bn/insidesearch/features/search/knowledge.html}} and Microsoft Satori\footnote{\url{https://searchengineland.com/library/bing/bing-satori}}.
	These knowledge graphs are successfully applied in many applications such as KG completion \cite{lin2015learning}, question answering \cite{dong2015question}, word embedding \cite{xu2014rc}, and text classification \cite{wang2017combining}.
	
	\begin{figure}[t]
		\centering
  		\includegraphics[width=.48\textwidth]{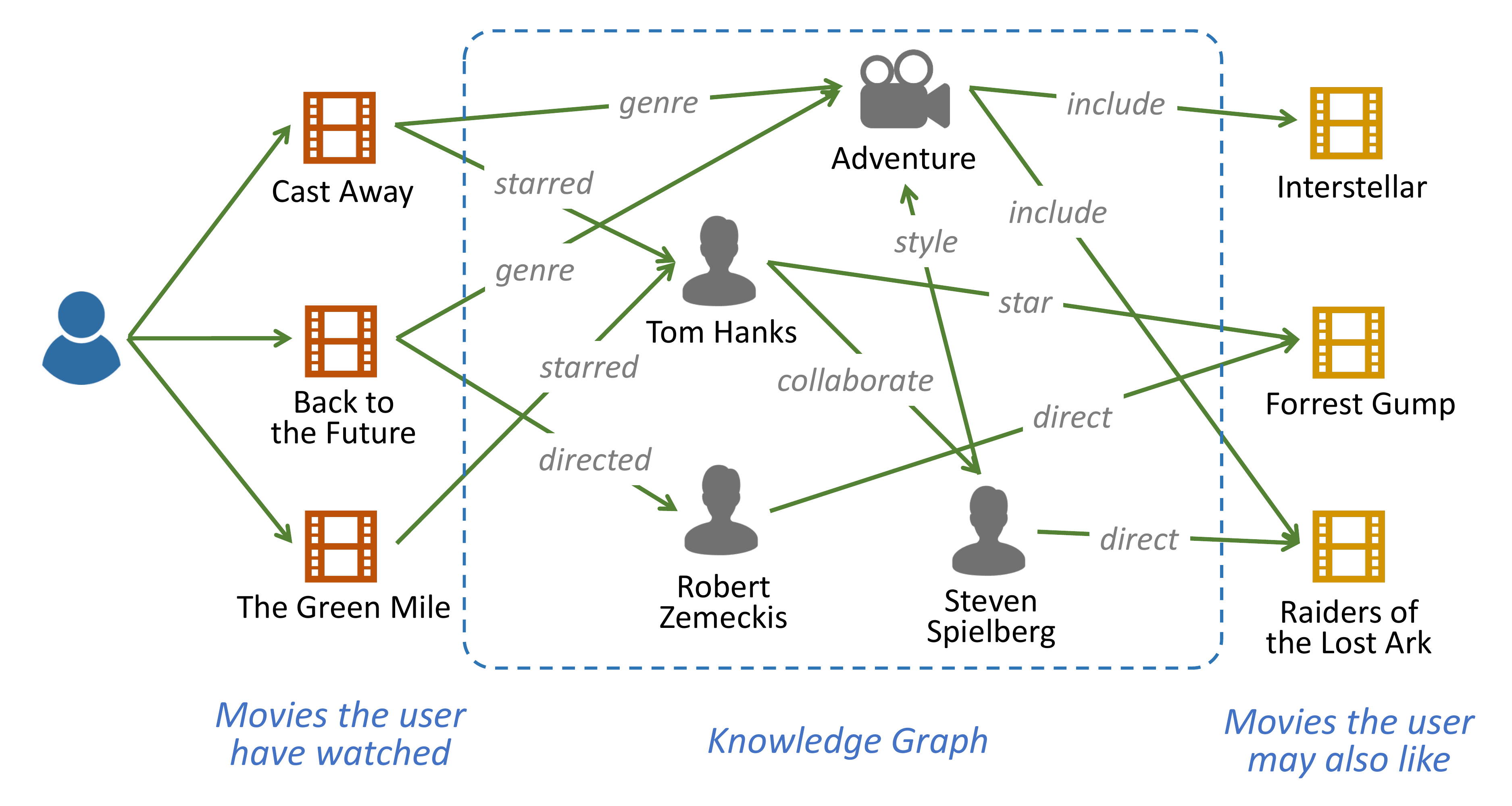}
  		\caption{Illustration of knowledge graph enhanced movie recommender systems. The knowledge graph provides fruitful facts and connections among items, which are useful for improving precision, diversity, and explainability of recommended results.}
  		\label{fig:kg}
	\end{figure}
	
	Inspired by the success of applying KG in a wide variety of tasks, researchers also tried to utilize KG to improve the performance of recommender systems.
	As shown in Figure \ref{fig:kg}, KG can benefit the recommendation from three aspects:
	(1) KG introduces semantic relatedness among items, which can help find their latent connections and improve the \textit{precision} of recommended items;
	(2) KG consists of relations with various types, which is helpful for extending a user's interests reasonably and increasing the \textit{diversity} of recommended items;
	(3) KG connects a user's historical records and the recommended ones, thereby bringing \textit{explainability} to recommender systems.
	In general, existing KG-aware recommendation can be classified into two categories:
	
	The first category is \textit{embedding-based methods} \cite{wang2018dkn,zhang2016collaborative,wang2018shine}, which pre-process a KG with \textit{knowledge graph embedding} (KGE) \cite{wang2017knowledge} algorithms and incorporates the learned entity embeddings into a recommendation framework.
	For example, Deep Knowledge-aware Network (DKN) \cite{wang2018dkn} treats entity embeddings and word embeddings as different channels, then designs a CNN framework to combine them together for news recommendation.
	Collaborative Knowledge base Embedding (CKE) \cite{zhang2016collaborative} combines a CF module with knowledge embedding, text embedding, and image embedding of items in a unified Bayesian framework.
	Signed Heterogeneous Information Network Embedding (SHINE) \cite{wang2018shine} designs deep autoencoders to embed sentiment networks, social networks and profile (knowledge) networks for celebrity recommendations.
	Embedding-based methods show high flexibility in utilizing KG to assist recommender systems, but the adopted KGE algorithms in these methods are usually more suitable for in-graph applications such as link prediction than for recommendation \cite{wang2017knowledge}, thus the learned entity embeddings are less intuitive and effective to characterize inter-item relations.
	
	The second category is \textit{path-based methods} \cite{yu2014personalized,zhao2017meta}, which explore the various patterns of connections among items in KG to provide additional guidance for recommendations.	
	For example, Personalized Entity Recommendation (PER) \cite{yu2014personalized} and Meta-Graph Based Recommendation \cite{zhao2017meta} treat KG as a heterogeneous information network (HIN), and extract meta-path/meta-graph based latent features to represent the connectivity between users and items along different types of relation paths/graphs.
	Path-based methods make use of KG in a more natural and intuitive way, but they rely heavily on manually designed meta-paths, which is hard to optimize in practice.
	Another concern is that it is impossible to design hand-crafted meta-paths in certain scenarios (e.g., news recommendation) where entities and relations are not within one domain.
	
	To address the limitations of existing methods, we propose \textit{RippleNet}, an end-to-end framework for knowledge-graph-aware recommendation.
	RippleNet is designed for click-through rate (CTR) prediction, which takes a user-item pair as input and outputs the probability of the user engaging (e.g., clicking, browsing) the item.
	The key idea behind RippleNet is \textit{preference propagation}:
	For each user, RippleNet treats his historical interests as a seed set in the KG, then extends the user's interests iteratively along KG links to discover his hierarchical potential interests with respect to a candidate item.
	We analogize preference propagation with actual ripples created by raindrops propagating on the water, in which multiple "ripples" superpose to form a resultant preference distribution of the user over the knowledge graph.
	The major difference between RippleNet and existing literature is that RippleNet combines the advantages of the above mentioned two types of methods:
	(1) RippleNet incorporates the KGE methods into recommendation naturally by preference propagation;
	(2) RippleNet can automatically discover possible paths from an item in a user's history to a candidate item, without any sort of hand-crafted design.
	
	Empirically, we apply RippleNet to three real-world scenarios of movie, book, and news recommendations.
	The experiment results show that RippleNet achieves AUC gains of $2.0\%$ to $40.6\%$, $2.5\%$ to $17.4\%$, and $2.6\%$ to $22.4\%$ in movie, book, and news recommendations, respectively, compared with state-of-the-art baselines for recommendation.
	We also find that RippleNet provides a new perspective of explainability for the recommended results in terms of the knowledge graph.
	
	In summary, our contributions in this paper are as follows:
	\begin{itemize}
		\item
			To the best of our knowledge, this is the first work to combine embedding-based and path-based methods in KG-aware recommendation.
		\item
			We propose RippleNet, an end-to-end framework utilizing KG to assist recommender systems.
			RippleNet automatically discovers users' hierarchical potential interests by iteratively propagating users' preferences in the KG.
		\item
			We conduct experiments on three real-world recommendation scenarios, and the results prove the efficacy of RippleNet over several state-of-the-art baselines.
	\end{itemize}

\section{Problem Formulation}\label{sec:problem_formulation}
	The knowledge-graph-aware recommendation problem is formulated as follows.
	In a typical recommender system, let $\mathcal U = \{u_1, u_2, ...\}$ and $\mathcal V = \{v_1, v_2, ...\}$ denote the sets of users and items, respectively.
	The user-item interaction matrix ${{\bf Y} = \{ y_{uv} | u \in \mathcal U, v \in \mathcal V \}}$ is defined according to users' implicit feedback, where
	\begin{equation}
		y_{uv} =
			\begin{cases}
				1, & $\rm{if interaction}$ \ (u, v) \ $\rm{is observed}$;\\
				0, & $\rm{otherwise}$.
			\end{cases}
	\end{equation}
	A value of 1 for $y_{uv}$ indicates there is an implicit interaction between user $u$ and item $v$, such as behaviors of clicking, watching, browsing, etc.
	In addition to the interaction matrix $\bf Y$, we also have a knowledge graph $\mathcal G$ available, which consists of massive entity-relation-entity triples $(h, r, t)$.
	Here $h \in \mathcal E$, $r \in \mathcal R$, and $t \in \mathcal E$ denote the head, relation, and tail of a knowledge triple, respectively, $\mathcal E$ and $\mathcal R$ denote the set of entities and relations in the KG.
	For example, the triple (\textit{Jurassic Park}, \textit{film.film.director}, \textit{Steven Spielberg}) states the fact that Steven Spielberg is the director of the film "Jurassic Park".
	In many recommendation scenarios, an item $v \in \mathcal V$ may associate with one or more entities in $\mathcal G$.
	For example, the movie "Jurassic Park" is linked with its namesake in KG, while news with title "France's Baby Panda Makes Public Debut" is linked with entities "France" and "panda".
	
	Given interaction matrix $\bf Y$ as well as knowledge graph $\mathcal G$, we aim to predict whether user $u$ has potential interest in item $v$ with which he has had no interaction before.
	Our goal is to learn a prediction function ${\hat y}_{uv} = \mathcal F(u, v ; \Theta)$, where ${\hat y}_{uv}$ denotes the probability that user $u$ will click item $v$, and $\Theta$ denotes the model parameters of function $\mathcal F$.

\section{RippleNet}\label{sec:ripple_networks}
	In this section, we discuss the proposed RippleNet in detail.
	We also give some discussions on the model and introduce the related work.
	
	\subsection{Framework}
		\begin{figure*}[t]
			\centering
  			\includegraphics[width=0.95\textwidth]{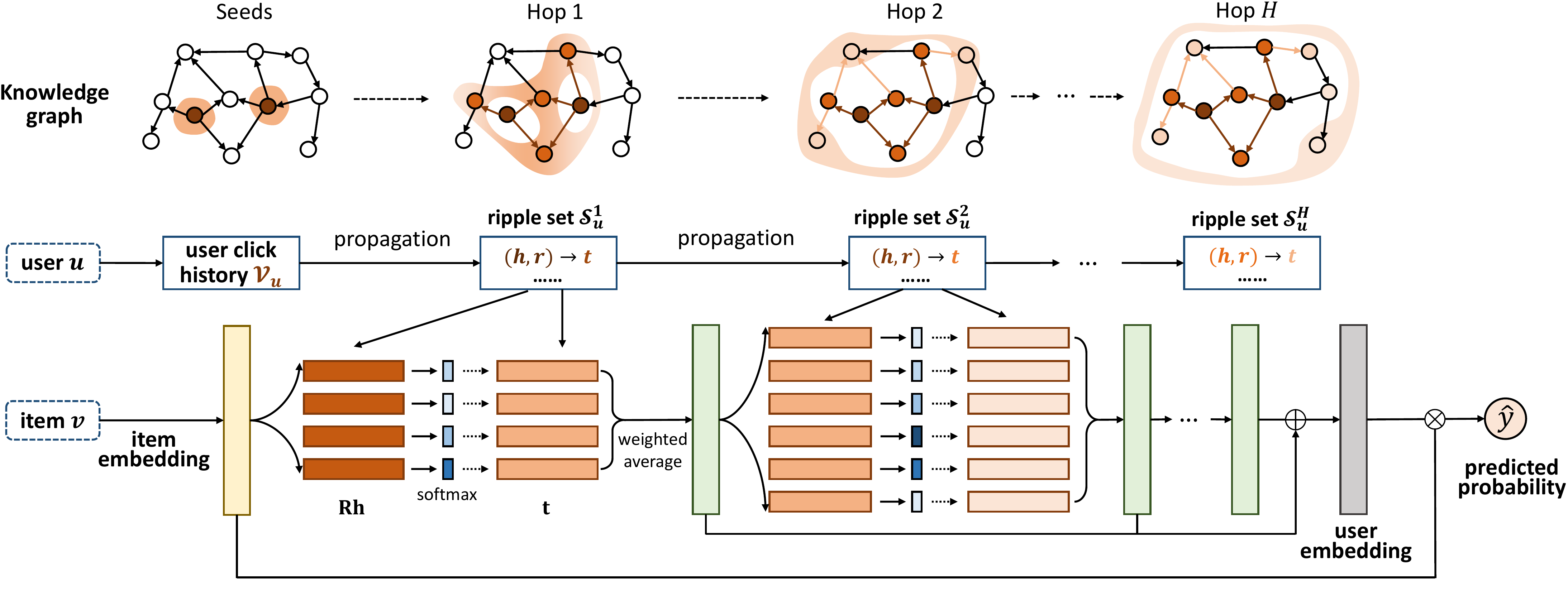}
  			\caption{The overall framework of the RippleNet. It takes one user and one item as input, and outputs the predicted probability that the user will click the item. The KGs in the upper part illustrate the corresponding ripple sets activated by the user's click history.}
  			\label{fig:framework}
		\end{figure*}	
		
		The framework of RippleNet is illustrated in Figure \ref{fig:framework}.
		RippleNet takes a user $u$ and an item $v$ as input, and outputs the predicted probability that user $u$ will click item $v$.
		For the input user $u$, his historical set of interests $\mathcal V_u$ is treated as seeds in the KG, then extended along links to form multiple ripple sets $\mathcal S_u^k$ ($k = 1, 2, ..., H$).
		A ripple set $\mathcal S_u^k$ is the set of knowledge triples that are $k$-hop(s) away from the seed set $\mathcal V_u$.
		These ripple sets are used to interact with the item embedding (the yellow block) iteratively for obtaining the responses of user $u$ with respect to item $v$ (the green blocks), which are then combined to form the final user embedding (the grey block).
		Lastly, we use the embeddings of user $u$ and item $v$ together to compute the predicted probability $\hat y_{uv}$.

	\subsection{Ripple Set}
	\label{sec:ripple_set}		
\begin{figure}[t]
			\centering
  			\includegraphics[width=.45\textwidth]{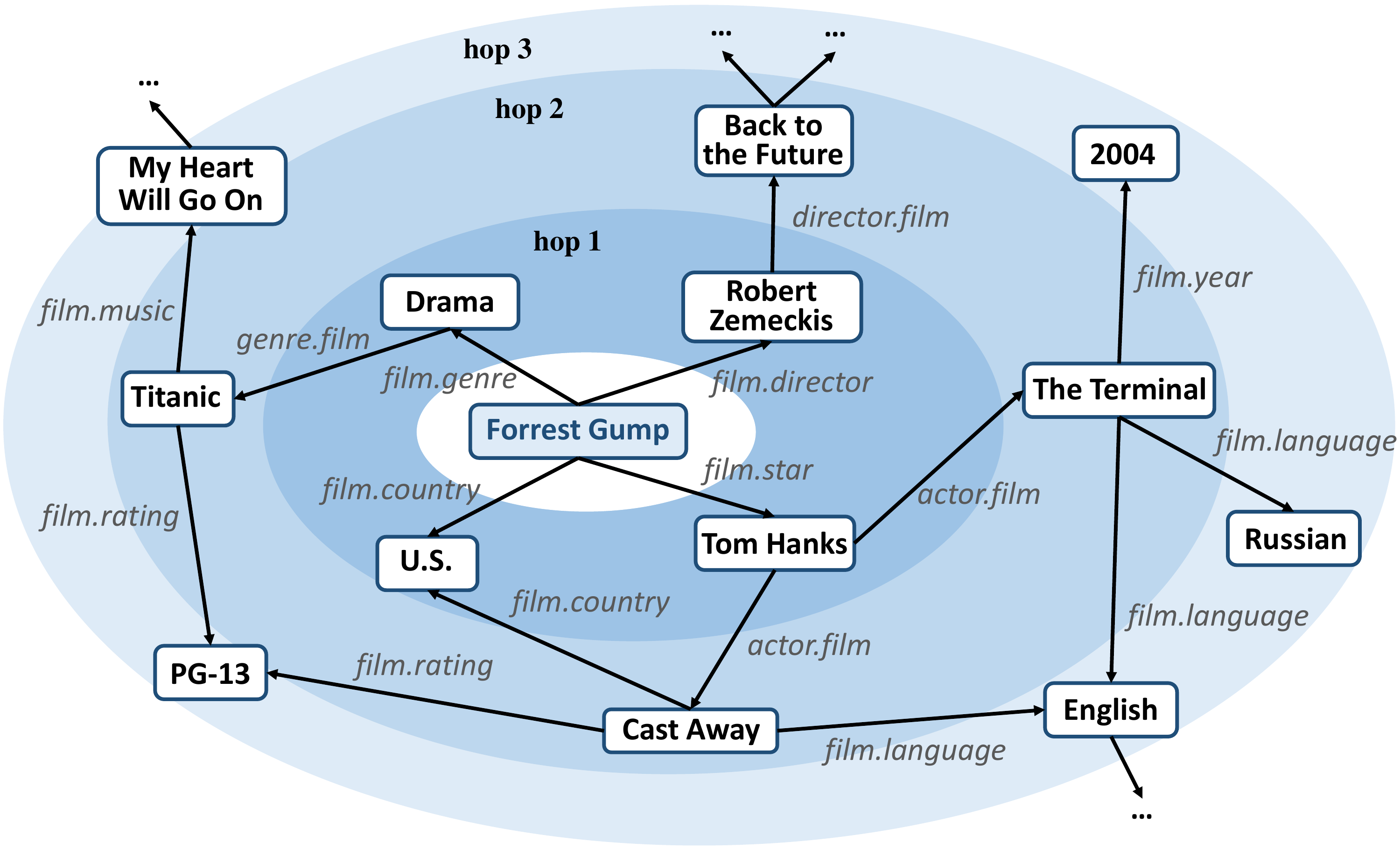}
  			\caption{Illustration of ripple sets of "Forrest Gump" in KG of movies. The concentric circles denotes the ripple sets with different hops. The fading blue indicates decreasing relatedness between the center and surrounding entities.Note that the ripple sets of different hops are not necessarily disjoint in practice.}
  			\label{fig:ripple_set}
		\end{figure}

		A knowledge graph usually contains fruitful facts and connections among entities.
		For example, as illustrated in Figure \ref{fig:ripple_set}, the film "Forrest Gump" is linked with "Robert Zemeckis" (director), "Tom Hanks" (star), "U.S." (country) and "Drama" (genre), while "Tom Hanks" is further linked with films "The Terminal" and "Cast Away" which he starred in.
		These complicated connections in KG provide us a deep and latent perspective to explore user preferences.
		For example, if a user has ever watched "Forrest Gump", he may possibly become a fan of Tom Hanks and be interested in "The Terminal" or "Cast Away".
		To characterize users' hierarchically extended preferences in terms of KG, in RippleNet, we recursively define the set of $k$-hop relevant entities for user $u$ as follows:
		
		\begin{definition}[relevant entity]
			Given interaction matrix $\bf Y$ and knowledge graph $\mathcal G$, the set of $k$-hop relevant entities for user $u$ is defined as
			\begin{equation}
			\label{eq:re}
			\mathcal E_u^k = \{ t \ | \ (h, r, t) \in \mathcal G \ \textrm{and} \ h \in \mathcal E_u^{k-1} \}, \quad k = 1, 2, ..., H,
		\end{equation}
		where $\mathcal E_u^0 = \mathcal V_u = \{ v \ | \ y_{uv} = 1 \}$ is the set of user's clicked items in the past, which can be seen as the seed set of user $u$ in KG.
		\end{definition}
		
		Relevant entities can be regarded as natural extensions of a user's historical interests with respect to the KG.
		Given the definition of relevant entities, we then define the $k$-hop ripple set of user $u$ as follows:
		
		\begin{definition}[ripple set]
			The $k$-hop ripple set of user $u$ is defined as the set of knowledge triples starting from $\mathcal E_u^{k-1}$:
		\begin{equation}
			\mathcal S_u^k = \{ (h, r, t) \ | \ (h, r, t) \in \mathcal G \ \textrm{and} \ h \in \mathcal E_u^{k-1} \}, \quad k = 1, 2, ..., H.
		\end{equation}
		\end{definition}
		
		The word "ripple" has two meanings:
		(1) Analogous to real ripples created by multiple raindrops, a user's potential interest in entities is activated by his historical preferences, then propagates along the links in KG layer by layer, from near to distant.
		We visualize the analogy by the concentric circles illustrated in Figure \ref{fig:ripple_set}.
		(2) The strength of a user's potential preferences in ripple sets weakens with the increase of the hop number $k$, which is similar to the gradually attenuated amplitude of real ripples.
		The fading blue in Figure \ref{fig:ripple_set} shows the decreasing relatedness between the center and surrounding entities.
		
		One concern about ripple sets is their sizes may get too large with the increase of hop number $k$.
		To address the concern, note that:
		(1) A large number of entities in a real KG are \textit{sink entities}, meaning they only have incoming links but no outgoing links, such as "2004" and "PG-13" in Figure \ref{fig:ripple_set}.
		(2) In specific recommendation scenarios such as movie or book recommendations, relations can be limited to scenario-related categories to reduce the size of ripple sets and improve relevance among entities.
		For example, in Figure \ref{fig:ripple_set}, all relations are movie-related and contain the word "film" in their names.
		(3) The number of maximal hop $H$ is usually not too large in practice, since entities that are too distant from a user's history may bring more noise than positive signals.
		We will discuss the choice of $H$ in the experiments part.
		(4) In RippleNet, we can sample a fixed-size set of neighbors instead of using a full ripple set to further reduce the computation overhead.
		Designing such samplers is an important direction of future work, especially the non-uniform samplers for better capturing user's hierarchical potential interests.
		
	\subsection{Preference Propagation}
		Traditional CF-based methods and their variants \cite{koren2008factorization,wang2017joint} learn latent representations of users and items, then predict unknown ratings by directly applying a specific function to their representations such as inner product.
		In RippleNet, to model the interactions between users and items in a more fine-grained way, we propose a preference propagation technique to explore users' potential interests in his ripple sets.
		
		As shown in Figure \ref{fig:framework}, each item $v$ is associated with an item embedding ${\bf v} \in {\mathbb R}^d$, where $d$ is the dimension of embeddings.
		Item embedding can incorporate one-hot ID \cite{koren2008factorization}, attributes \cite{wang2018shine}, bag-of-words \cite{wang2018dkn} or context information \cite{sun2017collaborative} of an item, based on the application scenario.
		Given the item embedding ${\bf v}$ and the 1-hop ripple set $\mathcal S_u^1$ of user $u$, each triple $(h_i, r_i, t_i)$ in $\mathcal S_u^1$ is assigned a relevance probability by comparing item $v$ to the head $h_i$ and the relation $r_i$ in this triple:
		\begin{equation}
		\label{eq:key_addressing}
			p_i = {\rm{softmax}} \left( {\bf v}^{\rm T} {\bf R}_i {\bf h}_i \right) = \frac{\exp \left( {\bf v}^{\rm T} {\bf R}_i {\bf h}_i \right)}{\sum_{(h, r, t) \in \mathcal S_u^1} \exp \left( {\bf v}^{\rm T} {\bf R} {\bf h} \right)},
		\end{equation}
		where ${\bf R}_i \in \mathbb R^{d \times d}$ and ${\bf h}_i \in \mathbb R^d$ are the embeddings of relation $r_i$ and head $h_i$, respectively.
		The relevance probability $p_i$ can be regarded as the similarity of item ${\bf v}$ and the entity ${\bf h}_i$ measured in the space of relation ${\bf R}_i$.
		Note that it is necessary to take the embedding matrix ${\bf R}_i$ into consideration when calculating the relevance of item ${\bf v}$ and entity ${\bf h}_i$, since an item-entity pair may have different similarities when measured by different relations.
		For example, "Forrest Gump" and "Cast Away" are highly similar when considering their directors or stars, but have less in common if measured by genre or writer.
		
		After obtaining the relevance probabilities, we take the sum of tails in $\mathcal S_u^1$ weighted by the corresponding relevance probabilities, and the vector ${\bf{o}}_u^1$ is returned:
		\begin{equation}
		\label{eq:value_reading}
			{\bf{o}}_u^1 = \sum\nolimits_{(h_i, r_i, t_i) \in \mathcal S_u^1} p_i {\bf t}_i,
		\end{equation}
		where ${\bf t}_i \in \mathbb R^d$ is the embedding of tail $t_i$.
		Vector ${\bf{o}}_u^1$ can be seen as the 1-order response of user $u$'s click history $\mathcal V_u$ with respect to item $v$.
		This is similar to item-based CF methods \cite{koren2008factorization,wang2018dkn}, in which a user is represented by his related items rather than a independent feature vector to reduce the size of parameters.
		Through the operations in Eq. (\ref{eq:key_addressing}) and Eq. (\ref{eq:value_reading}), a user's interests are transferred from his history set $\mathcal V_u$ to  the set of his 1-hop relevant entities $\mathcal E_u^1$ along the links in $\mathcal S_u^1$, which is called \textit{preference propagation} in RippleNet.
		
		Note that by replacing $\bf v$ with ${\bf{o}}_u^1$ in Eq. (\ref{eq:key_addressing}), we can repeat the procedure of preference propagation to obtain user $u$'s 2-order response ${\bf{o}}_u^2$, and the procedure can be performed iteratively on user $u$'s ripple sets $\mathcal S_u^i$ for $i = 1, ..., H$.
		Therefore, a user's preference is propagated up to $H$ hops away from his click history, and we observe multiple responses of user $u$ with different orders: ${\bf{o}}_u^1, {\bf{o}}_u^2, ..., {\bf{o}}_u^H$.
		The embedding of user $u$ with respect to item $v$ is calculated by combining the responses of all orders:
		\begin{equation}
			{\bf u} = {\bf{o}}_u^1 + {\bf{o}}_u^2 + ... + {\bf{o}}_u^H,
		\end{equation}
		Note that though the user response of last hop ${\bf{o}}_u^H$ contains all the information from previous hops theoretically, it is still necessary to incorporate ${\bf{o}}_u^k$ of small hops $k$ in calculating user embedding since they may be diluted in ${\bf{o}}_u^H$.
		Finally, the user embedding and item embedding are combined to output the predicted clicking probability:
		\begin{equation}
		\label{eq:inner}
			\hat y_{uv} = \sigma ({\bf u}^{\rm T} {\bf v}),
		\end{equation}
		where $\sigma(x) = \frac{1}{1 + \exp(-x)}$ is the sigmoid function.

	\subsection{Learning Algorithm}
		In RippleNet, we intend to maximize the following posterior probability of model parameters $\Theta$ after observing the knowledge graph $\mathcal G$ and the matrix of implicit feedback $\bf Y$:
		\begin{equation}
			\max p (\Theta | \mathcal G, {\bf Y}),
		\end{equation}
		where $\Theta$ includes the embeddings of all entities, relations and items.
		This is equivalent to maximizing
		\begin{equation}
		\label{eq:pp}
			p (\Theta | \mathcal G, {\bf Y}) = \frac{p (\Theta, \mathcal G, {\bf Y})}{p (\mathcal G, {\bf Y})} \propto p (\Theta) \cdot p (\mathcal G | \Theta) \cdot p ({\bf Y} | \Theta, \mathcal G)
		\end{equation}
		according to Bayes' theorem.
		In Eq. (\ref{eq:pp}), the first term $p (\Theta)$ measures the priori probability of model parameters $\Theta$.
		Following \cite{zhang2016collaborative}, we set $p (\Theta)$ as Gaussian distribution with zero mean and a diagonal covariance matrix:
		\begin{equation}
			p (\Theta) = \mathcal N({\bf 0}, \ \lambda_1^{-1} {\bf I}).
		\end{equation}
		The second item in Eq. (\ref{eq:pp}) is the likelihood function of the observed knowledge graph $\mathcal G$ given $\Theta$.
		Recently, researchers have proposed a great many knowledge graph embedding methods, including translational distance models \cite{bordes2013translating,lin2015learning} and semantic matching models \cite{nickel2016holographic,liu2017analogical} (We will continue the discussion on KGE methods in Section \ref{sec:kge}).
		In RippleNet, we use a three-way tensor factorization method to define the likelihood function for KGE:
		\begin{equation}
		\label{eq:kge}
			\begin{split}
				p (\mathcal G | \Theta) &= \prod\nolimits_{(h, r, t) \in \mathcal E \times \mathcal R \times \mathcal E} p \big( (h, r, t) | \Theta \big)\\
				&= \prod\nolimits_{(h, r, t) \in \mathcal E \times \mathcal R \times \mathcal E} \mathcal N \big( I_{h, r, t} - {\bf h}^{\rm T} {\bf R} {\bf t}, \ \lambda_2^{-1} \big),
			\end{split}
		\end{equation}
		where the indicator $I_{h, r, t} $ equals $1$ if $(h, r, t) \in \mathcal G$ and is $0$ otherwise.
		Based on the definition in Eq. (\ref{eq:kge}), the scoring functions of entity-entity pairs in KGE and item-entity pairs in preference propagation can thus be unified under the same calculation model.
		The last term in Eq. (\ref{eq:pp}) is the likelihood function of the observed implicit feedback given $\Theta$ and the KG, which is defined as the product of Bernouli distributions:
		\begin{equation}
			p ({\bf Y} | \Theta, \mathcal G) = \prod\nolimits_{(u, v) \in {\bf Y}} \sigma({\bf u}^{\rm T} {\bf v})^{y_{uv}} \cdot \big( 1-\sigma({\bf u}^{\rm T} {\bf v}) \big)^{1 - y_{uv}}
		\end{equation}
		based on Eq. (\ref{eq:re})--(\ref{eq:inner}).
		
		Taking the negative logarithm of Eq. (\ref{eq:pp}), we have the following loss function for RippleNet:
		\begin{equation}
		\label{eq:loss}
			\begin{split}
				\min \mathcal L =& -\log \big( p ({\bf Y} | \Theta, \mathcal G) \cdot p (\mathcal G | \Theta) \cdot p (\Theta) \big)\\
				=& \sum_{(u, v) \in {\bf Y}} - \Big( y_{uv} \log \sigma({\bf u}^{\rm T} {\bf v}) + (1 - y_{uv}) \log \big( 1-\sigma({\bf u}^{\rm T} {\bf v}) \big) \Big)\\
				&+ \frac{\lambda_2}{2} \sum_{r \in \mathcal R} \| {\bf I}_r - {\bf E}^{\rm T} {\bf R} {\bf E} \|_2^2 + \frac{\lambda_1}{2} \Big( \| {\bf V} \|_2^2 + \| {\bf E} \|_2^2 + \sum_{r \in \mathcal R} \| {\bf R} \|_2^2 \Big)\\
			\end{split}
		\end{equation}
		where $\bf V$ and $\bf E$ are the embedding matrices for all items and entities, respectively, ${\bf I}_r$ is the slice of the indicator tensor $\bf I$ in KG for relation $r$, and $\bf R$ is the embedding matrix of relation $r$.
		In Eq. (\ref{eq:loss}), The first term measures the cross-entropy loss between ground truth of interactions $\bf Y$ and predicted value by RippleNet, the second term measures the squared error between the ground truth of the KG ${\bf I}_r$ and the reconstructed indicator matrix ${\bf E}^{\rm T} {\bf R} {\bf E}$, and the third term is the regularizer for preventing over-fitting.
		
		\begin{algorithm}[t]
			\caption{Learning algorithm for RippleNet}
			\label{alg:rn}
			\begin{algorithmic}[1]
				\REQUIRE{Interaction matrix $\bf Y$, knowledge graph $\mathcal G$}
				\ENSURE{Prediction function $\mathcal F(u, v | \Theta)$}
				\STATE Initialize all parameters
				\STATE Calculate ripple sets $\{ \mathcal S_u^k \}_{k=1}^H$ for each user $u$;
				\FOR{number of training iteration}
				\STATE Sample minibatch of positive and negative interactions from $\bf Y$;
				\STATE Sample minibatch of true and false triples from $\mathcal G$;
				\STATE Calculate gradients $\partial \mathcal L / \partial \bf V$, $\partial \mathcal L / \partial \bf E$, $\{ \partial \mathcal L / \partial {\bf R} \}_{r \in \mathcal R}$, and $\{ \partial \mathcal L / \partial \alpha_i \}_{i=1}^H$ on the minibatch by back-propagation according to Eq. (4)-(13);
				\STATE Update $\bf V$, $\bf E$, $\{ \bf R \}_{r \in \mathcal R}$, and $\{ \alpha_i \}_{i=1}^H$ by gradient descent with learning rate $\eta$;
				\ENDFOR
				\RETURN $\mathcal F(u, v | \Theta)$
			\end{algorithmic}
		\end{algorithm}
		
		It is intractable to solve the above objection directly, therefore, we employ a stochastic gradient descent (SGD) algorithm to iteratively optimize the loss function.
		The learning algorithm of RippleNet is presented in Algorithm \ref{alg:rn}.
		In each training iteration, to make the computation more efficient, we randomly sample a minibatch of positive/negative interactions from $\bf Y$ and true/false triples from $\mathcal G$ following the negative sampling strategy in \cite{mikolov2013distributed}.
		Then we calculate the gradients of the loss $\mathcal L$ with respect to model parameters $\Theta$, and update all parameters by back-propagation based on the sampled minibatch.
		We will discuss the choice of hyper-parameters in the experiments section.

	\subsection{Discussion}
		\subsubsection{Explainability}
		\label{sec:explainability}
			Explainable recommender systems \cite{tintarev2007survey} aim to reveal why a user might like a particular item, which helps improve their acceptance or satisfaction of recommendations and increase trust in RS.
			The explanations are usually based on community tags \cite{vig2009tagsplanations}, social networks \cite{sharma2013social}, aspect \cite{bauman2017aspect}, and phrase sentiment \cite{zhang2014explicit}
			Since RippleNet explores users' interests based on the KG, it provides a new point of view of explainability by tracking the paths from a user's history to an item with high relevance probability (Eq. (\ref{eq:key_addressing})) in the KG.
			For example, a user's interest in film "Back to the Future" might be explained by the path "$user \xrightarrow{watched} Forrest \ Gump \xrightarrow{directed \ by} Robert \ Zemeckis \xrightarrow{directs} Back \ to \ the$ \textit{Future}", if the item "Back to the Future" is of high relevance probability with "Forrest Gump" and "Robert Zemeckis" in the user's $1$-hop and $2$-hop ripple set, respectively.
			Note that different from path-based methods \cite{yu2014personalized,zhao2017meta} where the patterns of path are manually designed, RippleNet automatically discovers the possible explanation paths according to relevance probability.
			We will further present a visualized example in the experiments section to intuitively demonstrate the explainability of RippleNet.

		\subsubsection{Ripple Superposition}
		\label{sec:ripple_superposition}
			A common phenomenon in RippleNet is that a user's ripple sets may be large in size, which dilutes his potential interests inevitably in preference propagation.
			However, we observe that relevant entities of different items in a user's click history often highly overlap.
			In other words, an entity could be reached by multiple paths in the KG starting from a user's click history.
			For example, "Saving Private Ryan" is connected to a user who has watched "The Terminal", "Jurassic Park" and "Braveheart" through actor "Tom Hanks", director "Steven Spielberg" and genre "War", respectively.
			These parallel paths greatly increase a user's interests in overlapped entities.
			We refer to the case as \textit{ripple superposition}, as it is analogous to the interference phenomenon in physics in which two waves superpose to form a resultant wave of greater amplitude in particular areas.
			The phenomenon of ripple superposition is illustrated in the second KG in Figure \ref{fig:framework}, where the darker red around the two lower middle entities indicates higher strength of the user's possible interests.
			We will also discuss ripple superposition in the experiments section.

	\subsection{Links to Existing Work}
		Here we continue our discussion on related work and make comparisons with existing techniques in a greater scope.
		
		\subsubsection{Attention Mechanism}
			The attention mechanism was originally proposed in image classification \cite{mnih2014recurrent} and machine translation \cite{bahdanau2014neural}, which aims to learn where to find the most relevant part of the input automatically as it is performing the task.
			The idea was soon transplanted to recommender systems \citep{wang2017dynamic,chen2017attentive,seo2017interpretable,zhou2017deep,wang2018dkn}.
			For example, DADM \cite{chen2017attentive} considers factors of specialty and date when assigning attention values to articles for recommendation;
			D-Attn \cite{seo2017interpretable} proposes an interpretable and dual attention-based CNN model that combines review text and ratings for product rating prediction;
			DKN \cite{wang2018dkn} uses an attention network to calculate the weight between a user's clicked item and a candidate item to dynamically aggregate the user's historical interests.
			RippleNet can be viewed as a special case of attention where tails are averaged weighted by similarities between their associated heads, tails, and certain item.
			The difference between our work and literature is that RippleNet designs a multi-level attention module based on knowledge triples for preference propagation.
		
		\subsubsection{Memory Networks}
			Memory networks \cite{weston2014memory,sukhbaatar2015end,miller2016key} is a recurrent attention model that utilizes an external memory module for question answering and language modeling.
			The iterative reading operations on the external memory enable memory networks to extract long-distance dependency in texts.
			Researchers have also proposed using memory networks in other tasks such as sentiment classification \cite{tai2015improved,li2017end} and recommendation \cite{huang2017mention,chen2018sequential}.
			Note that these works usually focus on entry-level or sentence-level memories, while our work addresses entity-level connections in the KG, which is more fine-grained and intuitive when performing multi-hop iterations.
			In addition, our work also incorporates a KGE term as a regularizer for more stable and effective learning.
			
		\subsubsection{Knowledge Graph Embedding}
		\label{sec:kge}
			RippleNet also connects to a large body of work in KGE methods\cite{bordes2013translating,wang2014knowledge,ji2015knowledge,lin2015learning,wang2018graphgan,nickel2016holographic,trouillon2016complex,yang2015embedding}.
			KGE intends to embed entities and relations in a KG into continuous vector spaces while preserving its inherent structure.
			Readers can refer to \cite{wang2017knowledge} for a more comprehensive survey.
			KGE methods are mainly classified into two categories:
			(1) Translational distance models, such as TransE \cite{bordes2013translating}, TransH \cite{wang2014knowledge}, TransD \cite{ji2015knowledge}, and TransR \cite{lin2015learning}, exploit distance-based scoring functions when learning representations of entities and relations.
			For example, TransE \cite{bordes2013translating} wants $\bf h + \bf r \approx \bf t$ when $(h, r, t)$ holds, where $\bf h$, $\bf r$ and $\bf t$ are the corresponding representation vector of $h$, $r$ and $t$.
			Therefore, TransE assumes the score function $f_r(h, t) = \| \bf h + \bf r - \bf t \|_2^2$ is low if $(h, r, t)$ holds, and high otherwise.
			(2) Semantic matching models, such as ANALOGY \cite{nickel2016holographic}, ComplEx \cite{trouillon2016complex}, and DisMult \cite{yang2015embedding}, measure plausibility of knowledge triples by matching latent semantics of entities and relations.
			For example, DisMult \cite{yang2015embedding} introduces a vector embedding ${\bf r} \in \mathbb R^d$ and requires ${\bf M}_r = \rm{diag}({\bf r})$.
			The scoring function is hence defined as $f_r(h, t) = {\bf h}^\top {\rm{diag}}({\bf r}){\bf t} = \sum_{i=1}^{d} [{\bf r}]_i \cdot [{\bf h}]_i \cdot [{\bf t}]_i$.
			Researchers also propose incorporating auxiliary information, such as entity types \cite{xie2016representation}, logic rules \cite{rocktaschel2015injecting}, and textual descriptions \cite{zhong2015aligning} to assist KGE.
			However, these methods are more suitable for in-graph applications such as link prediction or triple classification, according to their learning objectives.
			From this point of view, RippleNet can be seen as a specially designed KGE method that serves recommendation directly.

\section{Experiments}\label{section_experiments}
	In this section, we evaluate RippleNet on three real-world scenarios: movie, book, and news recommendations
	\footnote{Experiment code is provided at \url{https://github.com/hwwang55/RippleNet}.}.
	We first introduce the datasets, baselines, and experiment setup, then present the experiment results.
	We will also give a case study of visualization and discuss the choice of hyper-parameters in this section.

	\subsection{Datasets}
		We utilize the following three datasets in our experiments for movie, book, and news recommendation:
		\begin{itemize}
			\item
				MovieLens-1M\footnote{\url{https://grouplens.org/datasets/movielens/1m/}} is a widely used benchmark dataset in movie recommendations, which consists of approximately 1 million explicit ratings (ranging from 1 to 5) on the MovieLens website.
			\item
				Book-Crossing dataset\footnote{\url{http://www2.informatik.uni-freiburg.de/~cziegler/BX/}} contains 1,149,780 explicit ratings (ranging from 0 to 10) of books in the Book-Crossing community.
			\item
				Bing-News dataset contains 1,025,192 pieces of implicit feedback collected from the server logs of Bing News\footnote{\url{https://www.bing.com/news}} from October 16, 2016 to August 11, 2017.
				Each piece of news has a title and a snippet.
		\end{itemize}
		
		Since MovieLens-1M and Book-Crossing are explicit feedback data, we transform them into implicit feedback where each entry is marked with 1 indicating that the user has rated the item (the threshold of rating is 4 for MovieLens-1M, while no threshold is set for Book-Crossing due to its sparsity), and sample an unwatched set marked as 0 for each user, which is of equal size with the rated ones.
		For MovieLens-1M and Book-Crossing, we use the ID embeddings of users and items as raw input, while for Bing-News, we concatenate the ID embedding of a piece of news and the averaged word embedding of its title as raw input for the item, since news titles are typically much longer than names of movies or books, hence providing more useful information for recommendation.

		We use Microsoft Satori to construct the knowledge graph for each dataset.
		For MovieLens-1M and Book-Crossing, we first select a subset of triples from the whole KG whose relation name contains "movie" or "book" and the confidence level is greater than 0.9.
		Given the sub-KG, we collect IDs of all valid movies/books by matching their names with tail of triples (\textit{head, film.film.name, tail}) or \textit{(head, book.book.title, tail)}.
		For simplicity, items with no matched or multiple matched entities are excluded.
		We then match the IDs with the head and tail of all KG triples, select all well-matched triples from the sub-KG, and extend the set of entities iteratively up to four hops.
		The constructing process is similar for Bing-News except that: (1) we use entity linking tools to extract entities in news titles; (2) we do not impose restrictions on the names of relations since the entities in news titles are not within one particular domain.
		The basic statistics of the three datasets are presented in Table \ref{table:statistics}.

		\begin{table}[t]
			\centering
			\caption{Basic statistics of the three datasets.}
			\begin{tabular}{c|ccc}
				\hline
				& MovieLens-1M & Book-Crossing & Bing-News\\
				\hline
				\# users & 6,036 & 17,860 & 141,487\\
				\# items & 2,445 & 14,967 & 535,145\\
				\# interactions & 753,772 & 139,746 & 1,025,192\\
				\# 1-hop triples & 20,782 & 19,876 & 503,112\\
				\# 2-hop triples & 178,049 & 65,360 & 1,748,562\\
				\# 3-hop triples & 318,266 & 84,299 & 3,997,736\\
				\# 4-hop triples & 923,718 & 71,628 & 6,322,548\\
				\hline
			\end{tabular}
			\vspace{-0.07in}
			\label{table:statistics}
		\end{table}	
		
		\begin{table}[t]
			\centering
			\caption{Hyper-parameter settings for the three datasets.}
			\begin{tabular}{c|c}
				\hline
				MovieLens-1M & $d = 16$, $H = 2$, $\lambda_1 = 10^{-7}$, $\lambda_2 = 0.01$, $\eta = 0.02$\\
				Book-Crossing & $d = 4$, $H = 3$, $\lambda_1 = 10^{-5}$, $\lambda_2 = 0.01$, $\eta = 0.001$\\
				Bing-News & $d = 32$, $H = 3$, $\lambda_1 = 10^{-5}$, $\lambda_2 = 0.05$, $\eta = 0.005$\\
				\hline
			\end{tabular}
			\vspace{-0.07in}
			\label{table:hps}
		\end{table}	
		
	\subsection{Baselines}
		We compare the proposed RippleNet with the following state-of-the-art baselines:
		\begin{itemize}
			\item
				\textbf{CKE} \cite{zhang2016collaborative} combines CF with structural knowledge, textual knowledge, and visual knowledge in a unified framework for recommendation.
				We implement CKE as CF plus structural knowledge module in this paper.
			\item
				\textbf{SHINE} \cite{wang2018shine} designs deep autoencoders to embed a sentiment network, social network, and profile (knowledge) network for celebrity recommendation.
				Here we use autoencoders for user-item interaction and item profile to predict click probability.
			\item
				\textbf{DKN} \cite{wang2018dkn}	treats entity embedding and word embedding as multiple channels and combines them together in CNN for CTR prediction.
				In this paper, we use movie/book names and news titles as textual input for DKN.
			\item
				\textbf{PER} \cite{yu2014personalized} treats the KG as HIN and extracts meta-path based features to represent the connectivity between users and items.
				In this paper, we use all item-attribute-item features for PER (e.g., ``movie-director-movie").
			\item
				\textbf{LibFM} \cite{rendle2012factorization} is a widely used feature-based factorization model in CTR scenarios.
				We concatenate user ID, item ID, and the corresponding averaged entity embeddings learned from TransR \cite{lin2015learning} as input for LibFM.
			\item
				\textbf{Wide$\&$Deep} \cite{cheng2016wide} is a general deep model for recommendation combining a (wide) linear channel with a (deep) non-linear channel.
				Similar to LibFM, we use the embeddings of users, items, and entities to feed Wide$\&$Deep.
		\end{itemize}
		
	\begin{figure}[t]
			\centering
            \begin{subfigure}[b]{0.23\textwidth}
                \includegraphics[width=\textwidth]{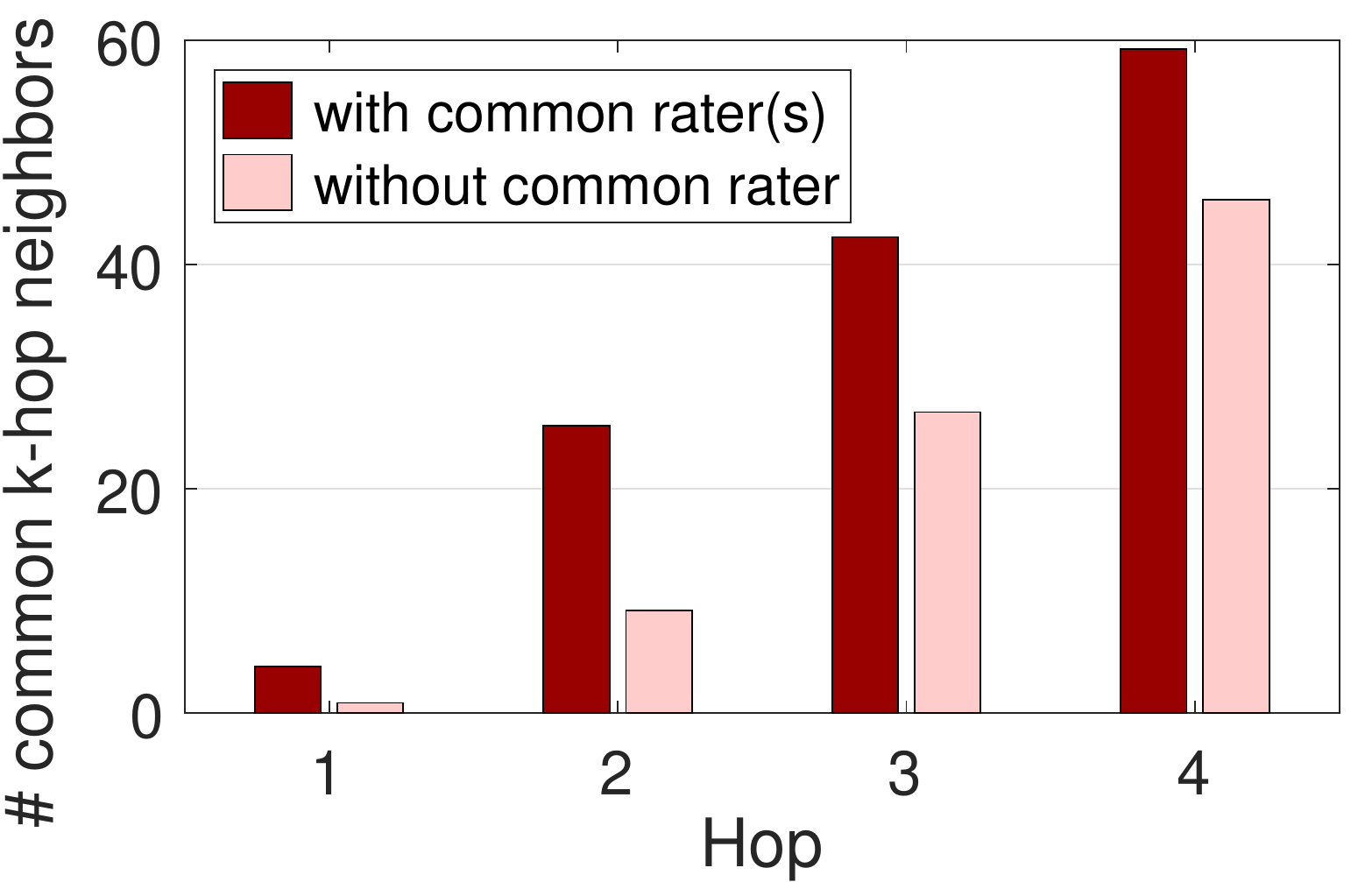}
                \caption{MovieLens-1M}
                \label{fig:es_1}
            \end{subfigure}
            \hfill
            \begin{subfigure}[b]{0.23\textwidth}
                \includegraphics[width=\textwidth]{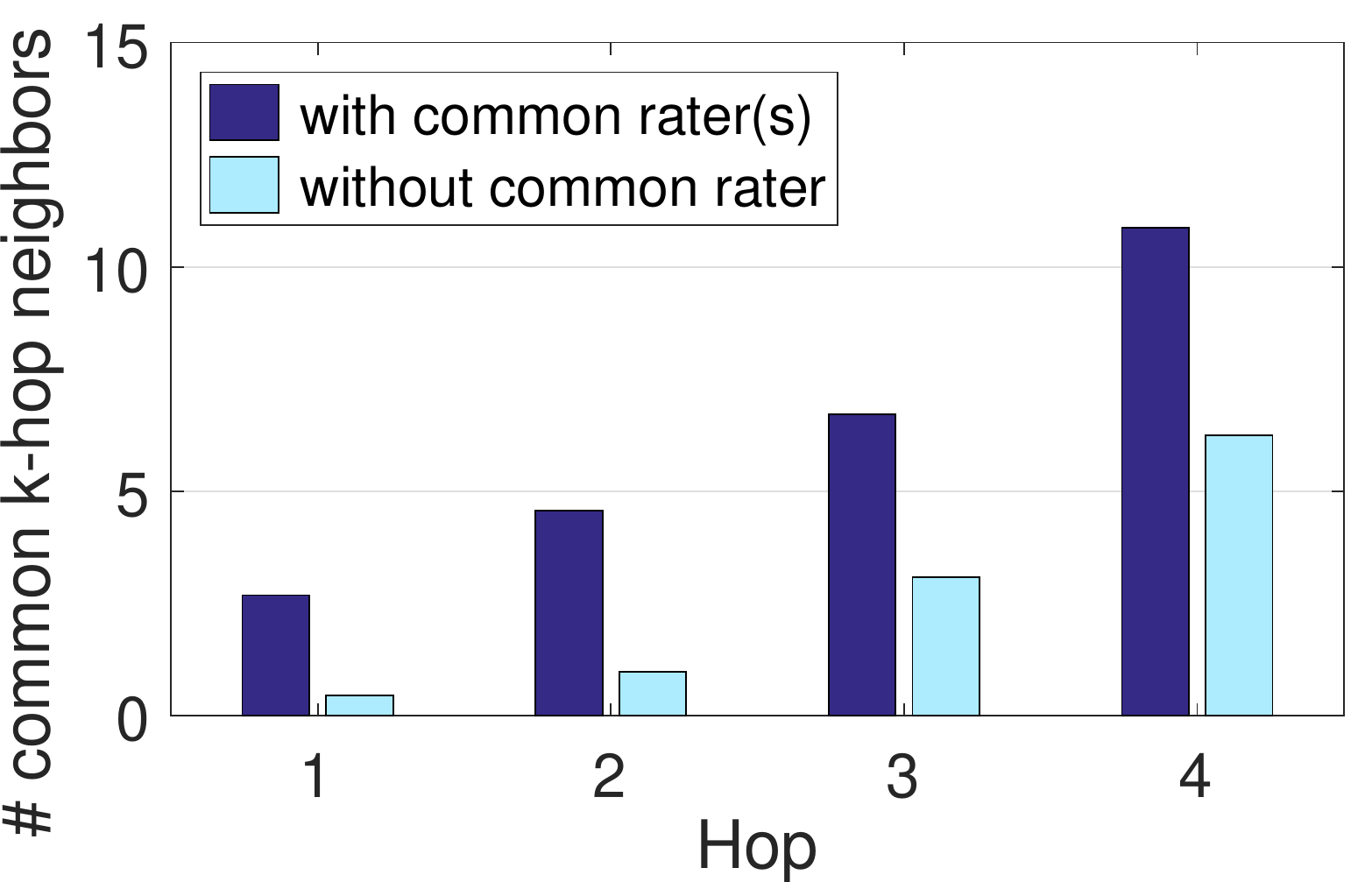}
                \caption{Book-Crossing}
                \label{fig:es_2}
            \end{subfigure}
            \hfill
            \begin{subfigure}[b]{0.23\textwidth}
                \includegraphics[width=\textwidth]{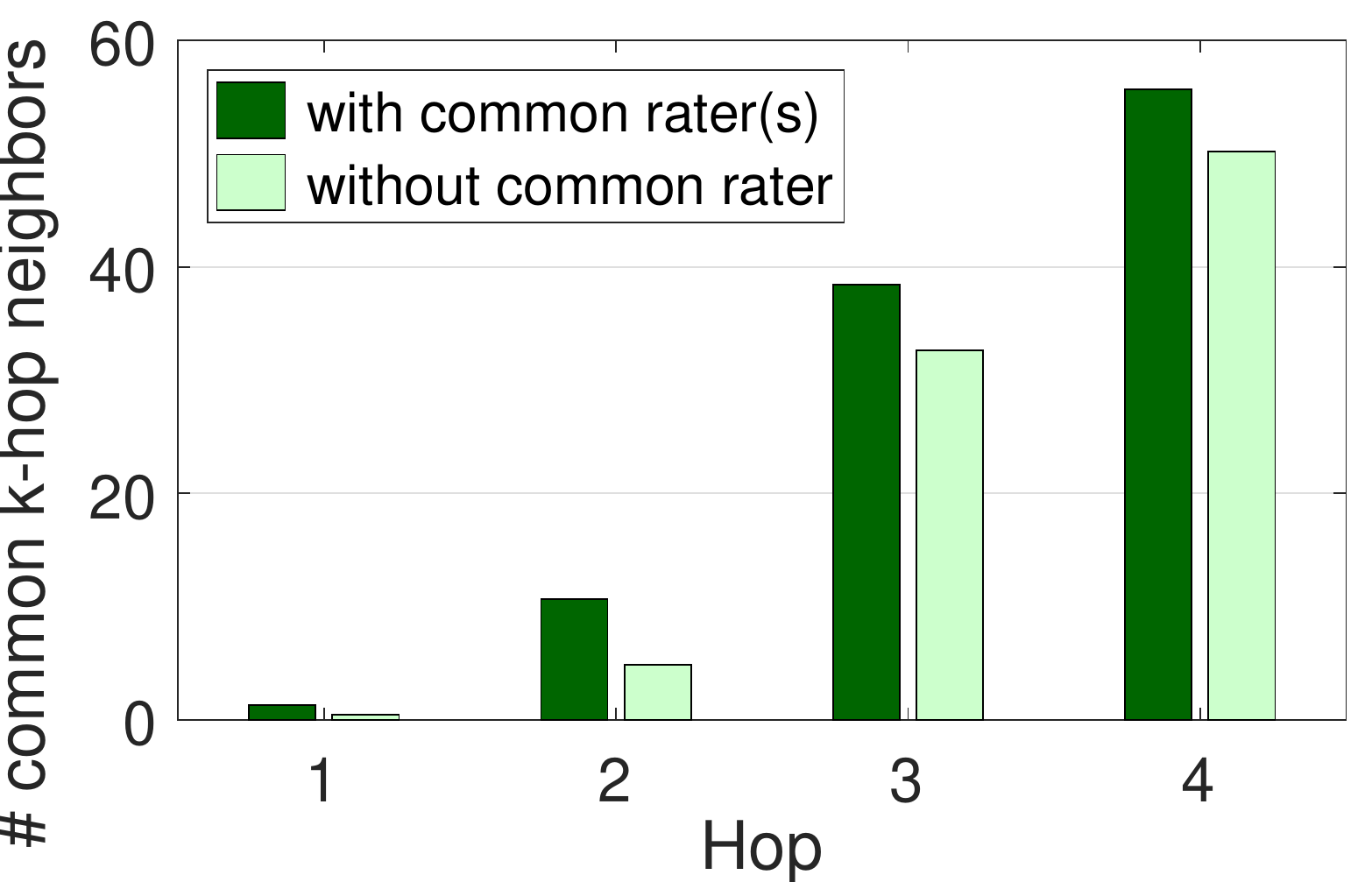}
                \caption{Bing-News}
                \label{fig:es_3}
            \end{subfigure}
            \hfill
            \begin{subfigure}[b]{0.23\textwidth}
                \includegraphics[width=\textwidth]{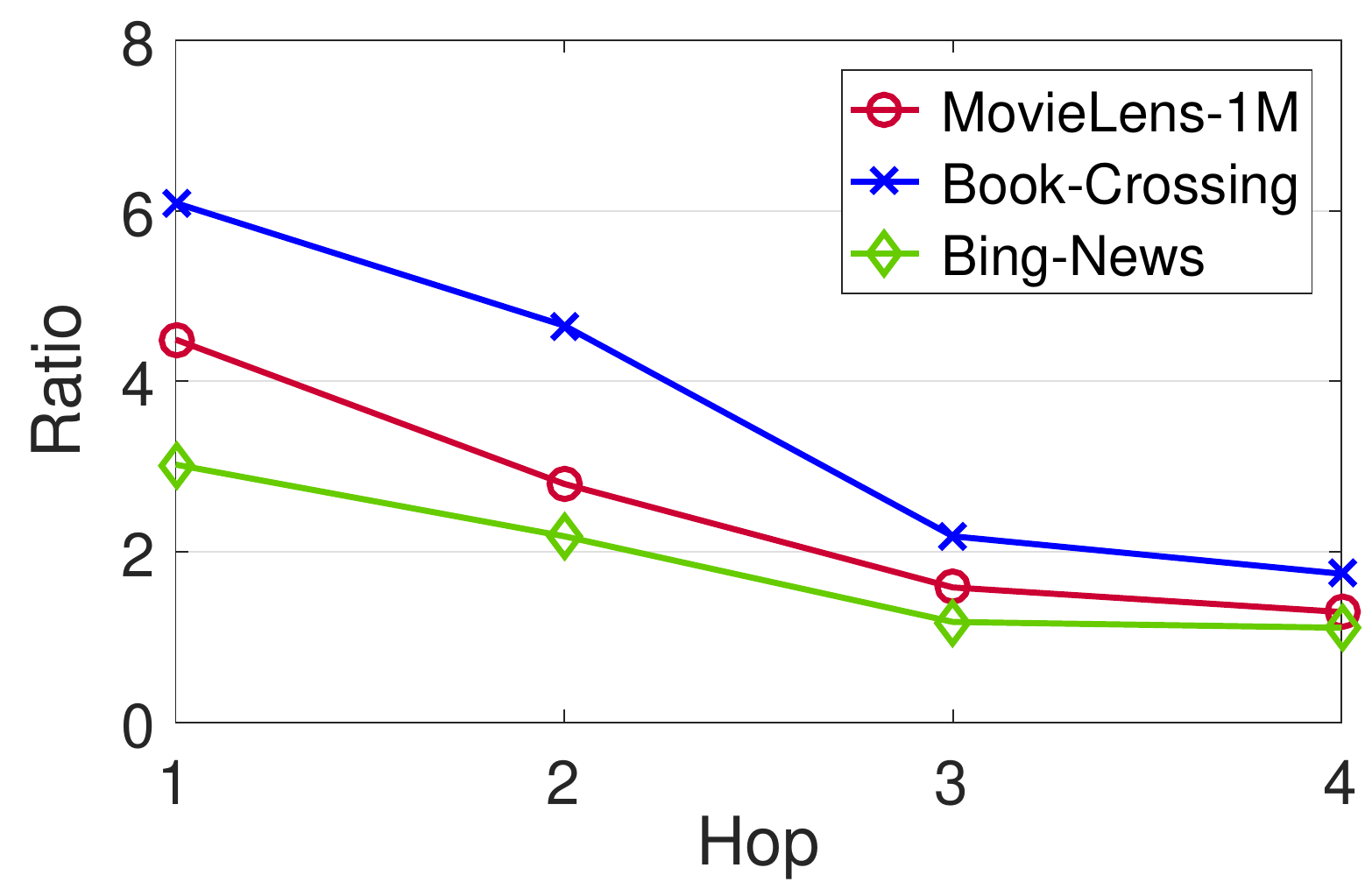}
                \caption{Ratio of two average numbers}
                \label{fig:es_4}
            \end{subfigure}
            \caption{The average number of $k$-hop neighbors that two items share in the KG w.r.t. whether they have common raters in (a) MovieLens-1M, (b) Book-Crossing, and (c) Bing-News datasets. (d) The ratio of the two average numbers with different hops.}
            \label{fig:case_study}
        \end{figure}

	\subsection{Experiment Setup}	
		In RippleNet, we set the hop number $H = 2$ for MovieLens-1M/Book-Crossing and $H = 3$ for Bing-News.
		A larger number of hops hardly improves performance but does incur heavier computational overhead according to experiment results. 
		The complete hyper-parameter settings are given in Table \ref{table:hps}, where $d$ denotes the dimension of embedding for items and the knowledge graph, and $\eta$ denotes the learning rate.
		The hyper-parameters are determined by optimizing $AUC$ on a validation set.
		For fair consideration, the latent dimensions of all compared baselines are set the same as in Table \ref{table:hps}, while other hyper-parameters of baselines are set based on grid search.
		
		For each dataset, the ratio of training, evaluation, and test set is $6:2:2$.
		Each experiment is repeated $5$ times, and the average performance is reported.
		We evaluate our method in two experiment scenarios:
		(1) In click-through rate (CTR) prediction, we apply the trained model to each piece of interactions in the test set and output the predicted click probability.
		We use $Accuracy$ and $AUC$ to evaluate the performance of CTR prediction.
		(2) In top-$K$ recommendation, we use the trained model to select $K$ items with highest predicted click probability for each user in the test set, and choose $Precision@K$, $Recall@K$, $F1@K$ to evaluate the recommended sets.

	\subsection{Empirical Study}
	\label{sec:es}        
        We conduct an empirical study to investigate the correlation between the average number of common neighbors of an item pair in the KG and whether they have common rater(s) in RS.
		For each dataset, we first randomly sample one million item pairs, then count the average number of $k$-hop neighbors that the two items share in the KG under the following two circumstances: (1) the two items have at least one common rater in RS; (2) the two items have no common rater in RS.
		The results are presented in Figures \ref{fig:es_1}, \ref{fig:es_2}, \ref{fig:es_3}, respectively, which clearly show that if two items have common rater(s) in RS, they likely share more common $k$-hop neighbors in the KG for fixed $k$.
		The above findings empirically demonstrate that \textit{the similarity of proximity structures of two items in the KG could assist in measuring their relatedness in RS}.
		In addition, we plot the ratio of the two average numbers with different hops (i.e., dividing the higher bar by its immediate lower bar for each hop number) in Figure \ref{fig:es_4}, from which we observe that the proximity structures of two items under the two circumstances become more similar with the increase of the hop number.
		This is because any two items are probable to share a large amount of $k$-hop neighbors in the KG for a large $k$, even if there is no direct similarity between them in reality.
		The result motivates us to find a moderate hop number in RippleNet to explore users' potential interests as far as possible while avoiding introducing too much noise.

	\subsection{Results}
		\begin{table}[t]
			\setlength{\abovecaptionskip}{3pt}
            \centering
            \caption{The results of $AUC$ and $Accuracy$ in CTR prediction.}
                \begin{tabular}{c|cccccc}
                    \hline
                    \multirow{2}{*}{Model} & \multicolumn{2}{c}{MovieLens-1M} & \multicolumn{2}{c}{Book-Crossing} & \multicolumn{2}{c}{Bing-News} \\
                    \cline{2-7}
                    & \textit{AUC} & \textit{ACC} & \textit{AUC} & \textit{ACC} & \textit{AUC} & \textit{ACC} \\
                    \hline
                    RippleNet* & \textbf{0.921} & \textbf{0.844} & \textbf{0.729} & \textbf{0.662} & \textbf{0.678} & \textbf{0.632} \\
                    CKE & 0.796 & 0.739 & 0.674 & 0.635 & 0.560 & 0.517 \\
                    SHINE & 0.778 & 0.732 & 0.668 & 0.631 & 0.554 & 0.537 \\
                    DKN & 0.655 & 0.589 & 0.621 & 0.598 & 0.661 & 0.604 \\
                    PER & 0.712 & 0.667 & 0.623 & 0.588 & - & - \\
                    LibFM & 0.892 & 0.812 & 0.685 & 0.639 & 0.644 & 0.588 \\
                    Wide$\&$Deep & 0.903 & 0.822 & 0.711 & 0.623 & 0.654 & 0.595 \\
                    \hline
				\end{tabular}
			\label{table:ctr}
			\footnotesize \flushleft{* Statistically significant improvements by unpaired two-sample $t$-test with $p=0.1$.}
		\end{table}
		
		\begin{figure*}[t]
			\centering
			\begin{subfigure}[b]{0.9\textwidth}
				\vspace{-0.1in}
                \includegraphics[width=\textwidth]{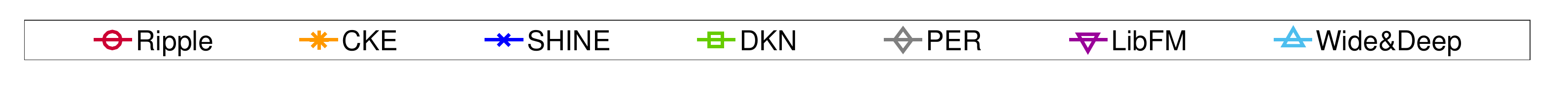}
                \vspace{-0.3in}
            \end{subfigure}
            \hfill
            \begin{subfigure}[b]{0.3\textwidth}
                \includegraphics[width=\textwidth]{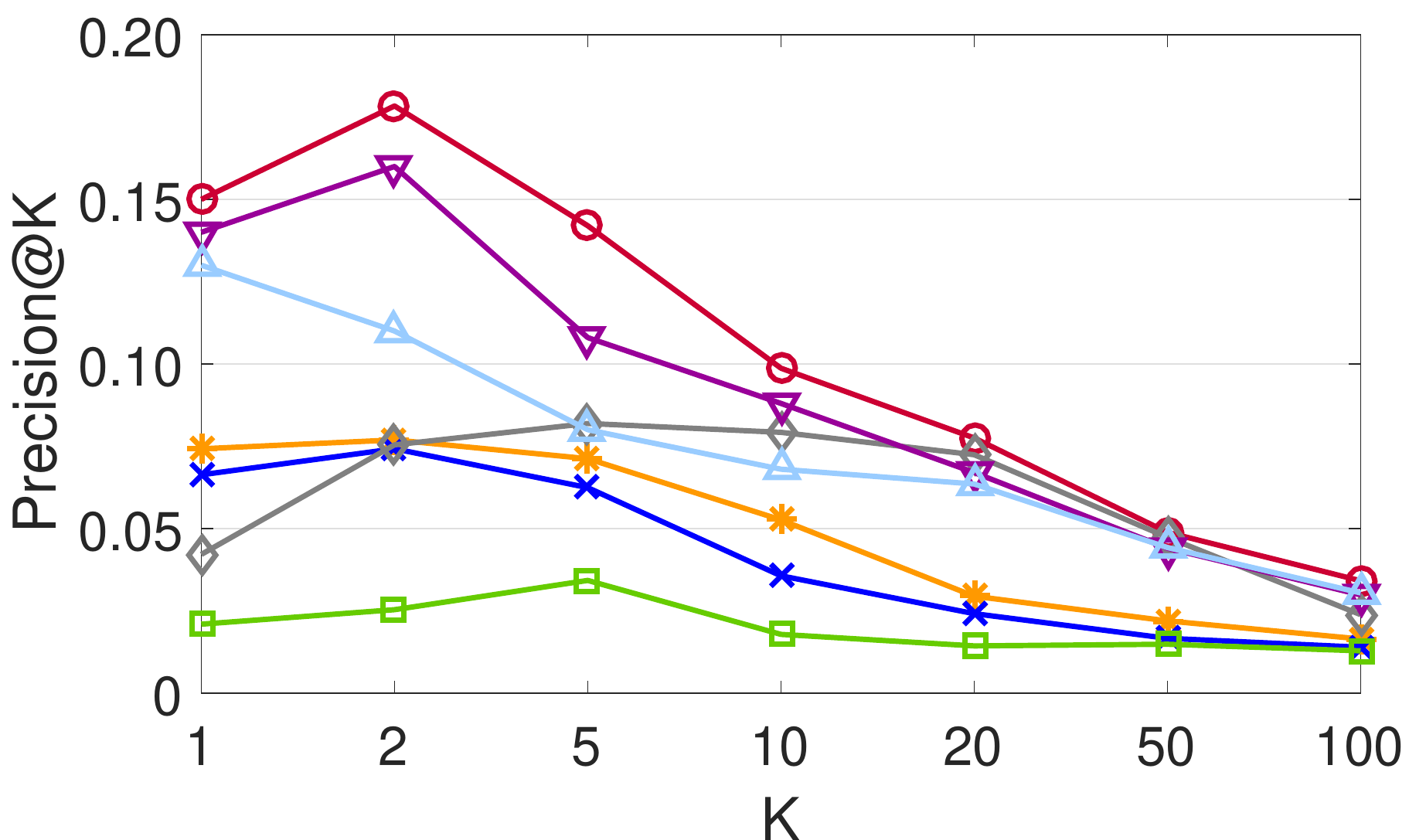}
                \caption{$Precision@K$}
            \end{subfigure}
            \hfill
            \begin{subfigure}[b]{0.3\textwidth}
                \includegraphics[width=\textwidth]{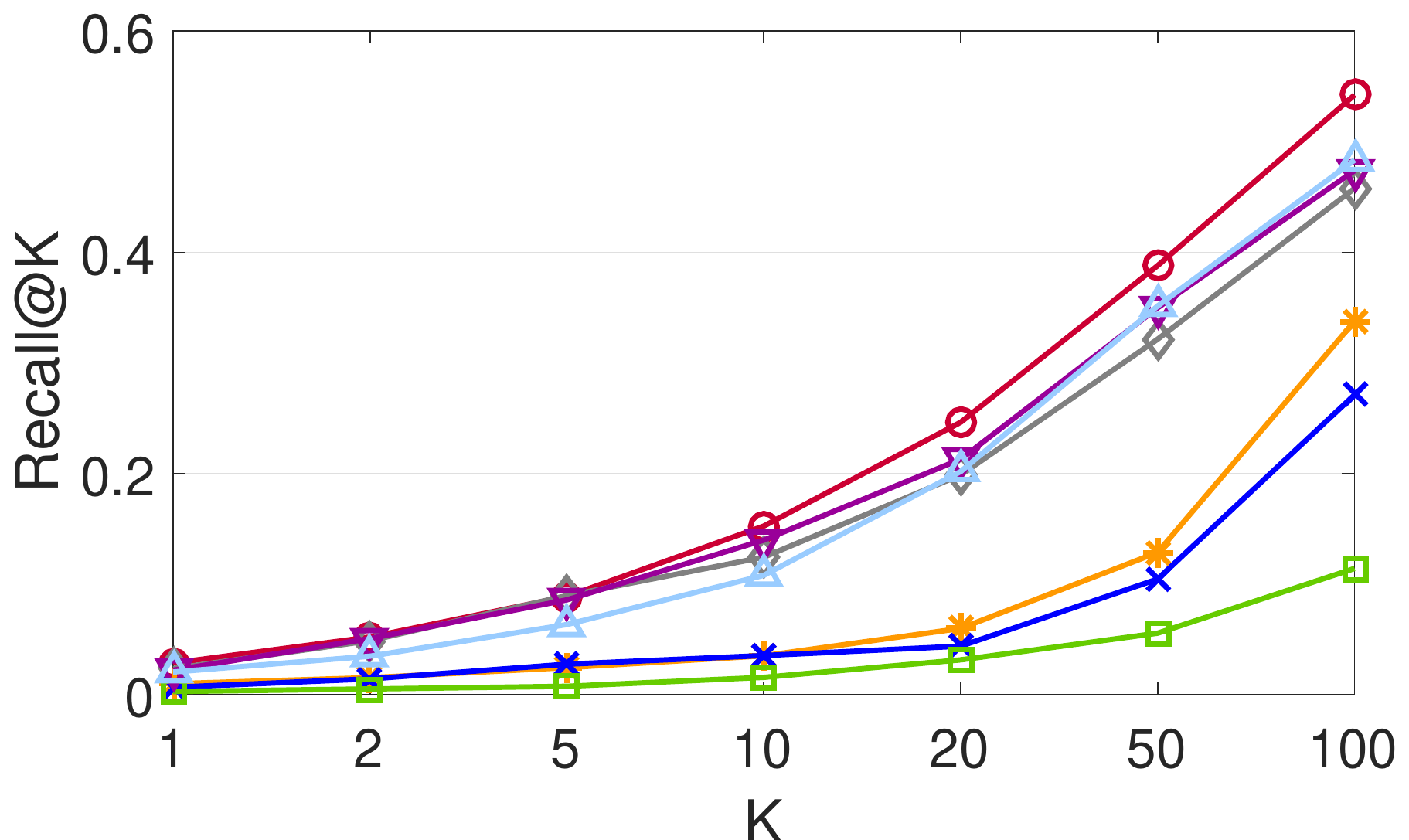}
                \caption{$Recall@K$}
            \end{subfigure}
            \hfill
            \begin{subfigure}[b]{0.3\textwidth}
                \includegraphics[width=\textwidth]{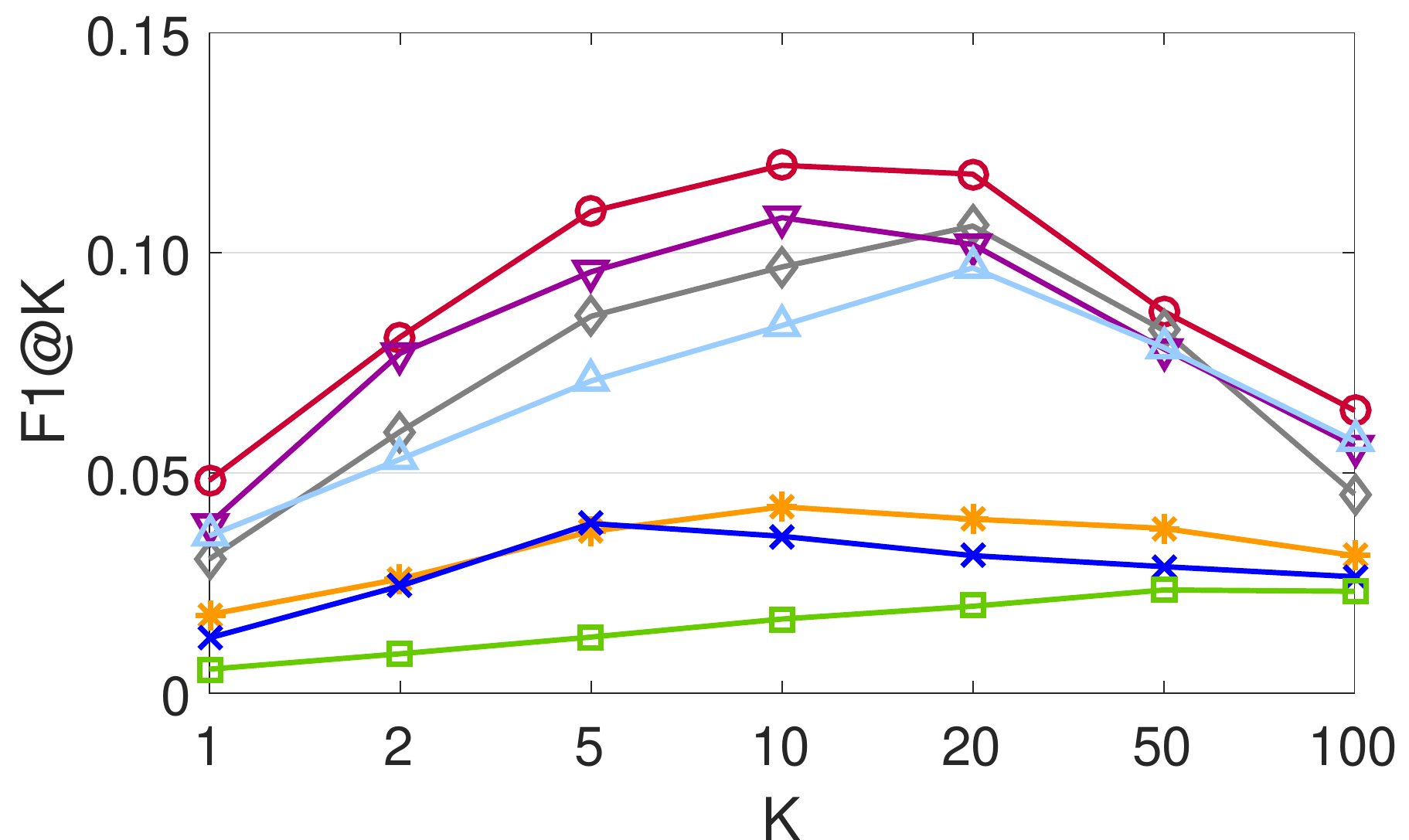}
                \caption{$F1@K$}
            \end{subfigure}
            \caption{$Precision@K$, $Recall@K$, and $F1@K$ in top-$K$ recommendation for MovieLens-1M.}
            \label{fig:topk_movie}
        \end{figure*}
        
        \begin{figure*}[t]
			\centering
            \begin{subfigure}[b]{0.3\textwidth}
                \includegraphics[width=\textwidth]{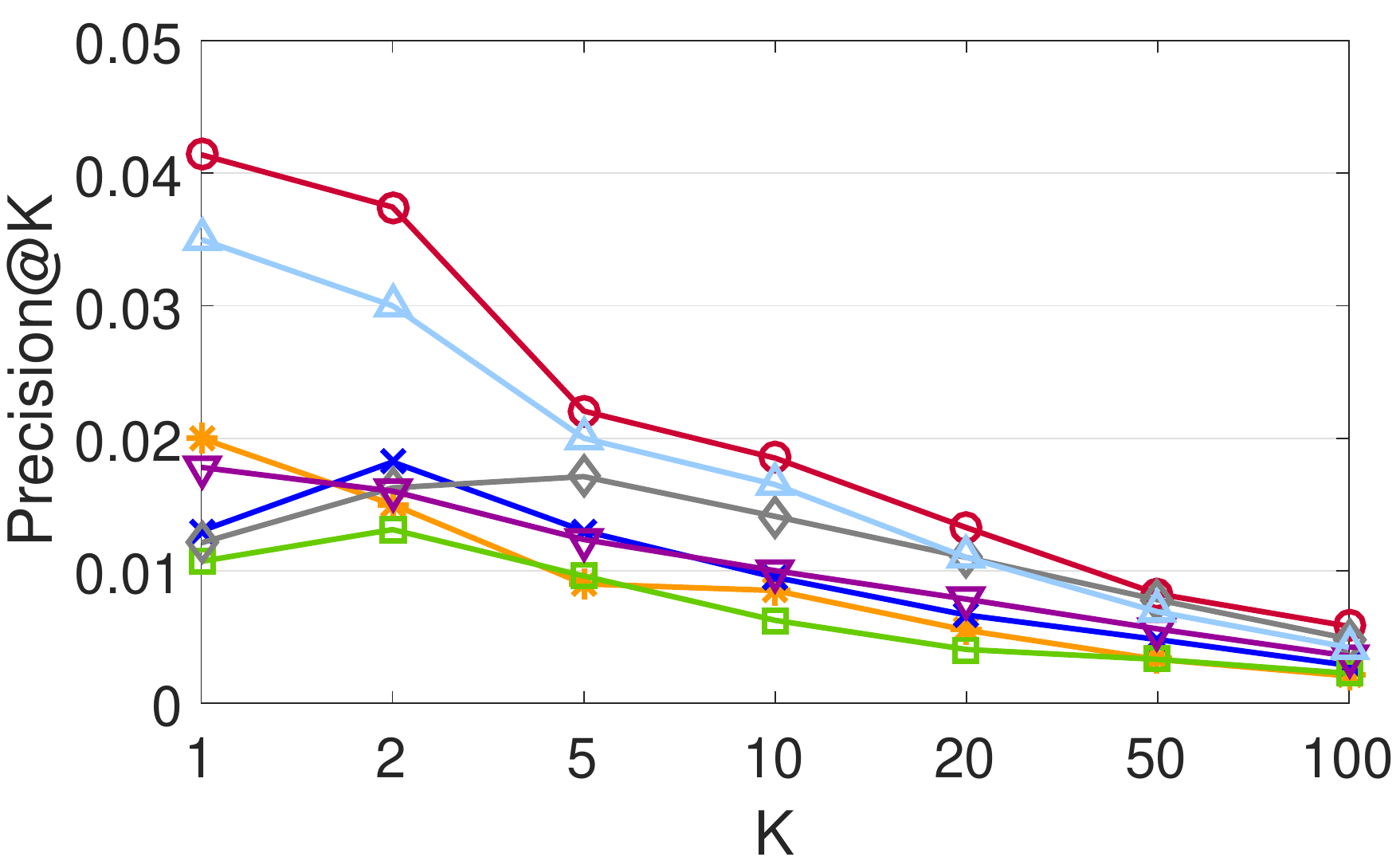}
                \caption{$Precision@K$}
            \end{subfigure}
            \hfill
            \begin{subfigure}[b]{0.3\textwidth}
                \includegraphics[width=\textwidth]{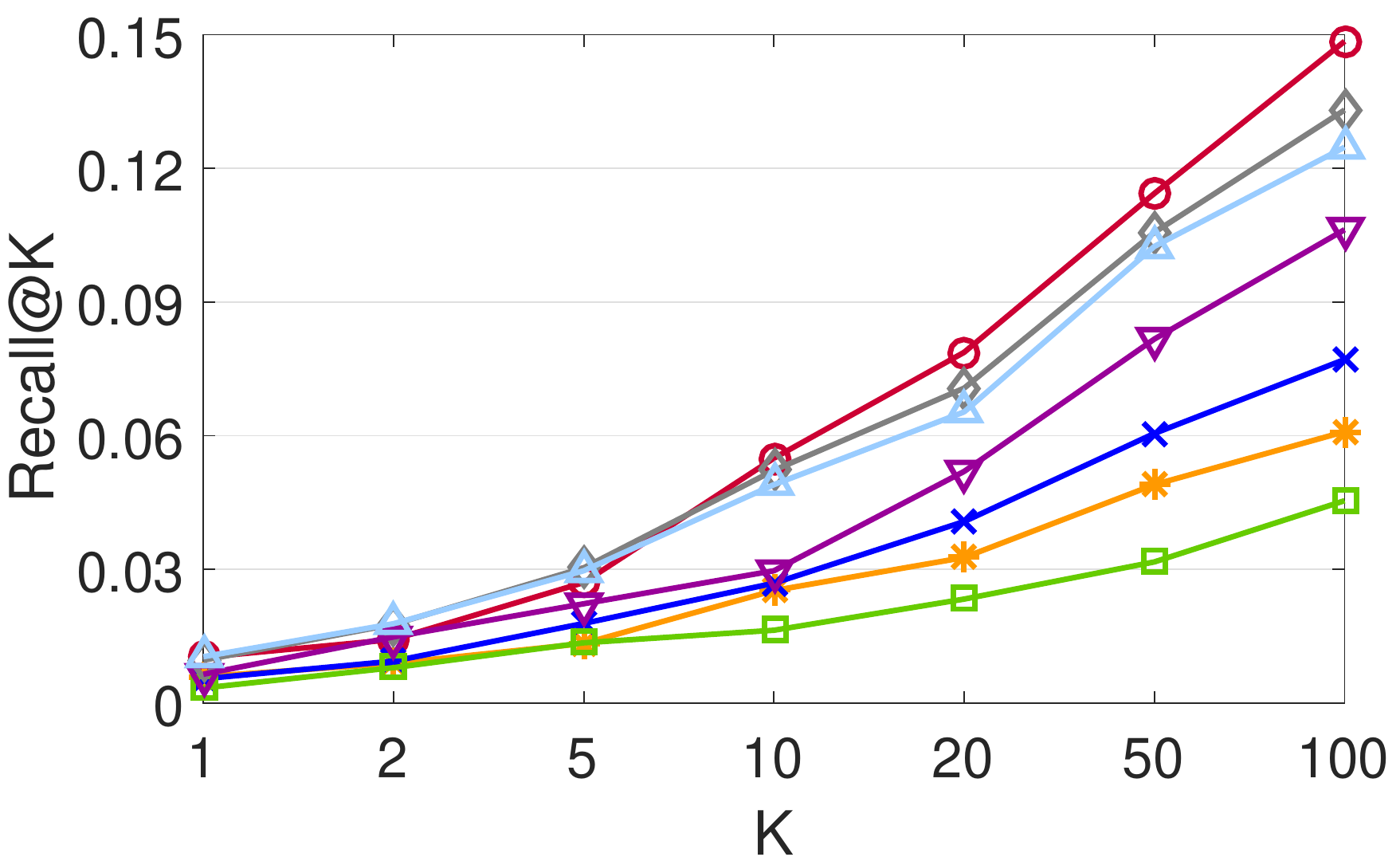}
                \caption{$Recall@K$}
            \end{subfigure}
            \hfill
            \begin{subfigure}[b]{0.3\textwidth}
                \includegraphics[width=\textwidth]{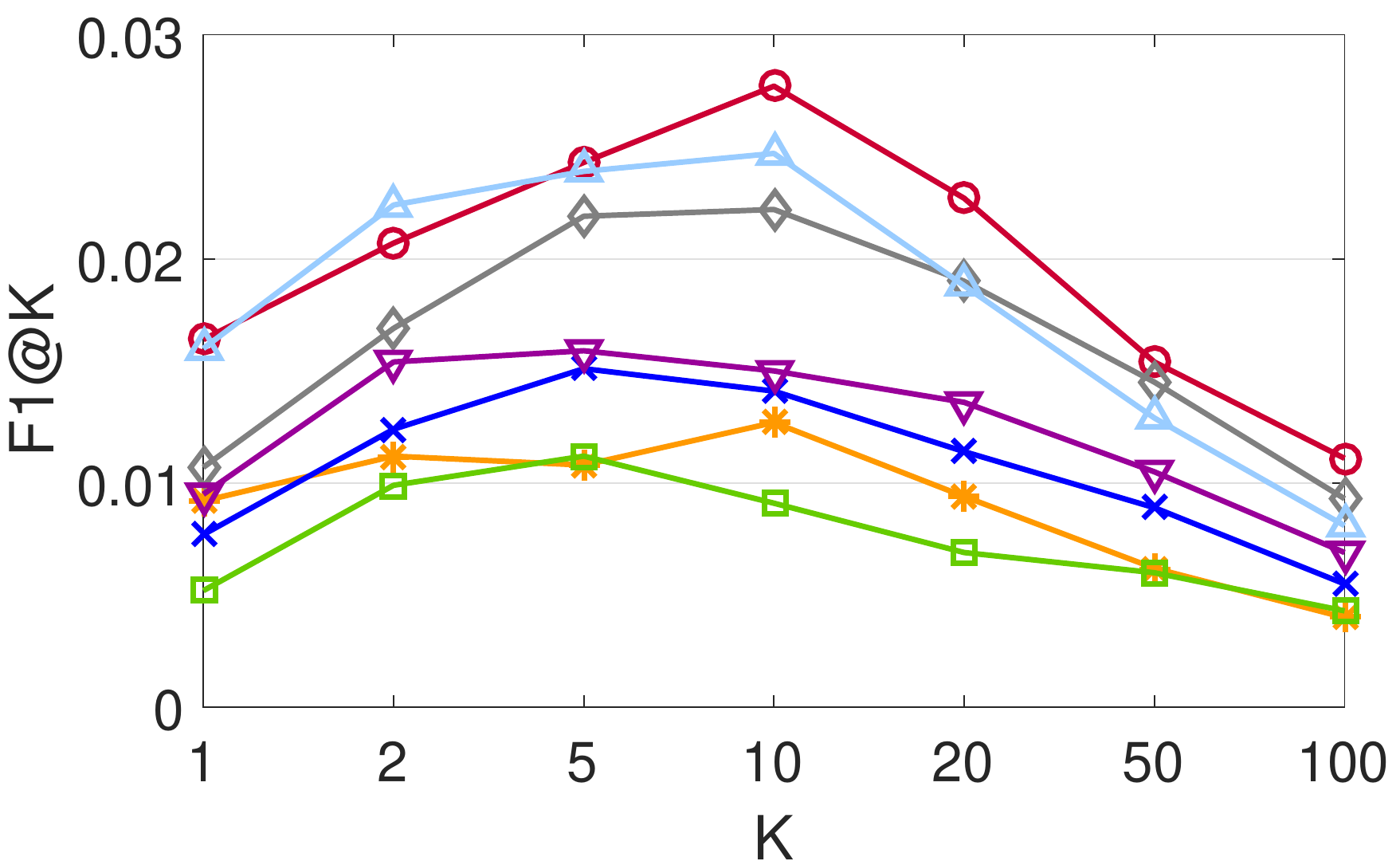}
                \caption{$F1@K$}
            \end{subfigure}
            \caption{$Precision@K$, $Recall@K$, and $F1@K$ in top-$K$ recommendation for Book-Crossing.}
            \label{fig:topk_book}
        \end{figure*}
        
        \begin{figure*}[t]
			\centering
            \begin{subfigure}[b]{0.3\textwidth}
                \includegraphics[width=\textwidth]{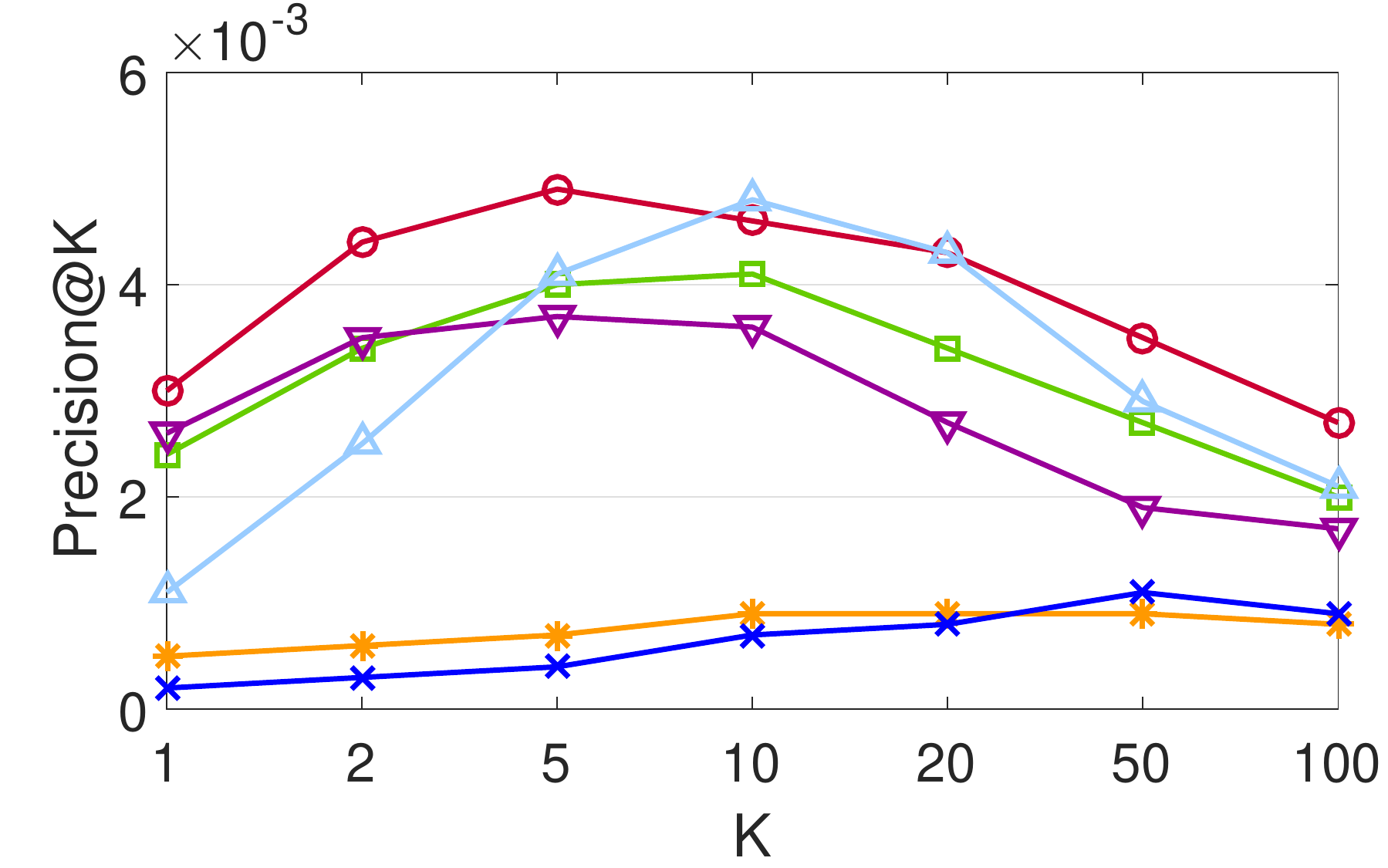}
                \caption{$Precision@K$}
            \end{subfigure}
            \hfill
            \begin{subfigure}[b]{0.3\textwidth}
                \includegraphics[width=\textwidth]{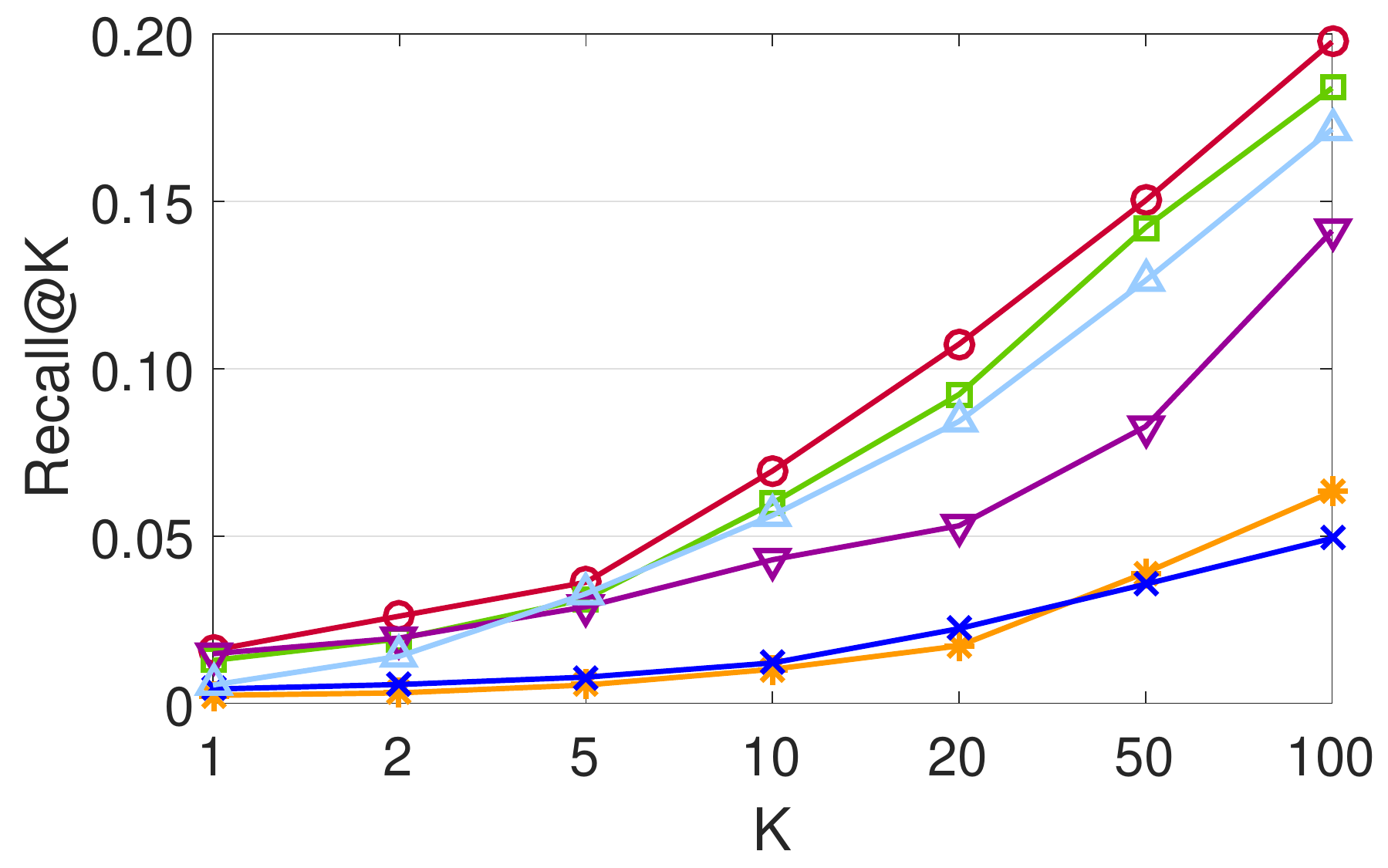}
                \caption{$Recall@K$}
            \end{subfigure}
            \hfill
            \begin{subfigure}[b]{0.3\textwidth}
                \includegraphics[width=\textwidth]{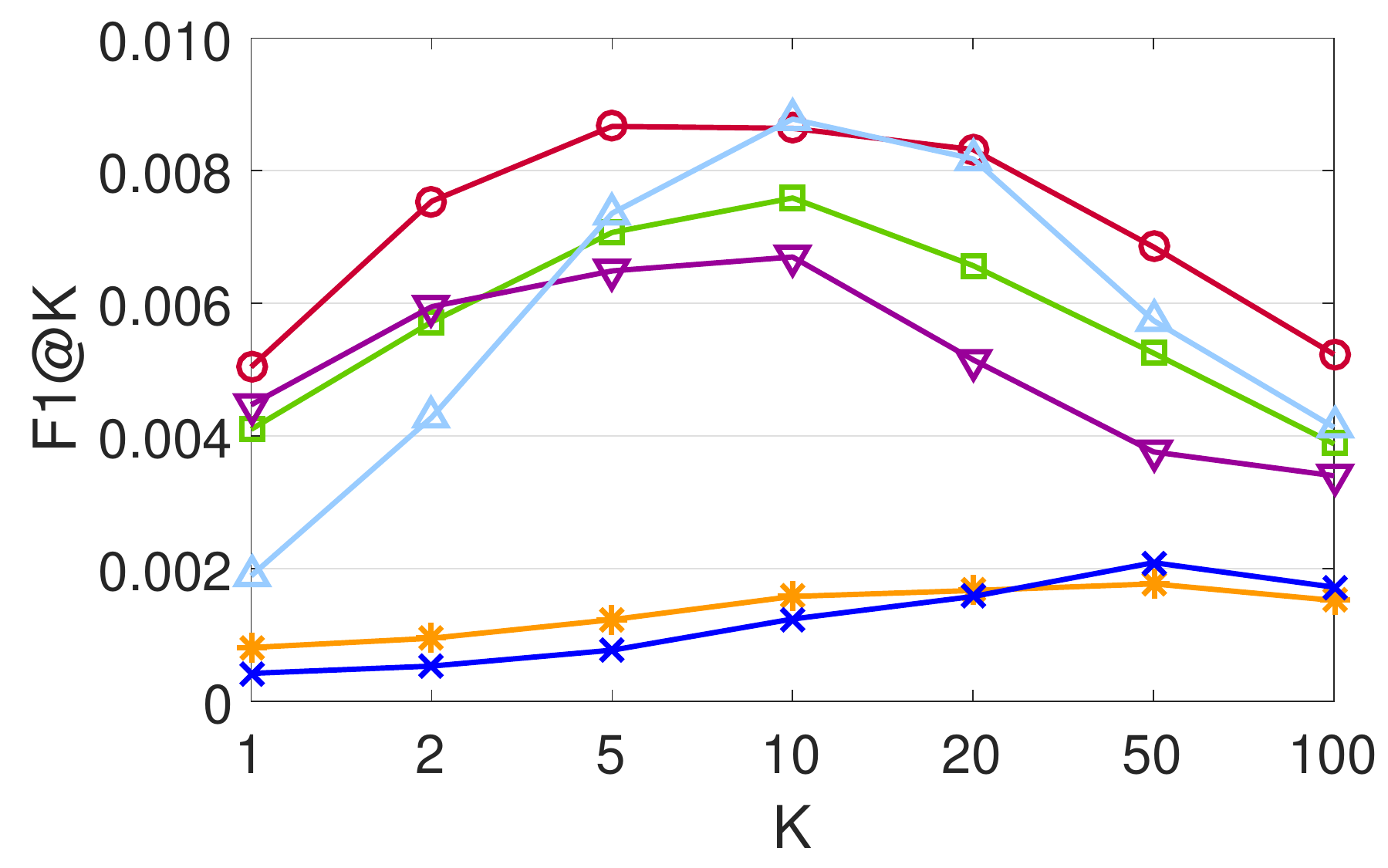}
                \caption{$F1@K$}
            \end{subfigure}
            \caption{$Precision@K$, $Recall@K$, and $F1@K$ in top-$K$ recommendation for Bing-News.}
            \label{fig:topk_news}
        \end{figure*}
        
        The results of all methods in CTR prediction and top-$K$ recommendation are presented in Table \ref{table:ctr} and Figures \ref{fig:topk_movie}, \ref{fig:topk_book}, \ref{fig:topk_news}, respectively.
        Several observations stand out:
        \begin{itemize}
        	\item
        		CKE performs comparably poorly than other baselines, which is probably because we only have structural knowledge available, without visual and textual input.
        	\item
        		SHINE performs better in movie and book recommendation than news.
        		This is because the 1-hop triples for news are too complicated when taken as profile input.
        	\item
        		DKN performs best in news recommendation compared with other baselines, but performs worst in movie and book recommendation.
        		This is because movie and book names are too short and ambiguous to provide useful information.
        	\item
        		PER performs unsatisfactorily on movie and book recommendation because the user-defined meta-paths can hardly be optimal.
        		In addition, it cannot be applied in news recommendation since the types of entities and relations involved in news are too complicated to pre-define meta-paths.
        	\item
        		As two generic recommendation tools, LibFM and Wide$\&$Deep achieve satisfactory performance, demonstrating that they can make well use of knowledge from KG into their algorithms.
        	\item
        		RippleNet performs best among all methods in the three datasets.
        		Specifically, RippleNet outperforms baselines by $2.0\%$ to $40.6\%$, $2.5\%$ to $17.4\%$, and $2.6\%$ to $22.4\%$ on $AUC$ in movie, book, and news recommendation, respectively.
        		RippleNet also achieves outstanding performance in top-$K$ recommendation as shown in Figures \ref{fig:topk_movie}, \ref{fig:topk_book}, and \ref{fig:topk_news}.
        		Note that the performance of top-$K$ recommendation is much lower for Bing-News because the number of news is significantly larger than movies and books.
        \end{itemize}
        
        \noindent \textbf{Size of ripple set in each hop}.
        We vary the size of a user's ripple set in each hop to further investigate the robustness of RippleNet.
        The results of $AUC$ on the three datasets are presented in Table \ref{table:kg_ratio}, from which we observe that with the increase of the size of ripple set, the performance of RippleNet is improved at first because a larger ripple set can encode more knowledge from the KG.
        But notice that the performance drops when the size is too large.
        In general, a size of $16$ or $32$ is enough for most datasets according to the experiment results.

        \begin{table}[t]
			\setlength{\abovecaptionskip}{3pt}
            \centering
            \caption{The results of $AUC$ w.r.t. different sizes of a user's ripple set.}
                \begin{tabular}{c|cccccc}
                    \hline
                    Size of ripple set & 2 & 4 & 8 & 16 & 32 & 64\\
                    \hline
                    MovieLens-1M & 0.903 & 0.908 & 0.911 & 0.918 & \textbf{0.920} & 0.919 \\
                    Book-Crossing & 0.694 & 0.696 & 0.708 & \textbf{0.726} & 0.706 & 0.711 \\
                    Bing-News & 0.659 & 0.672 & 0.670 & 0.673 & \textbf{0.678} & 0.671 \\
                    \hline
				\end{tabular}
			\label{table:kg_ratio}
		\end{table}
        
        \noindent \textbf{Hop number}.
        We also vary the maximal hop number $H$ to see how performance changes in RippleNet.
        The results are shown in Table \ref{table:hop_number}, which shows that the best performance is achieved when $H$ is $2$ or $3$.
        We attribute the phenomenon to the trade-off between the positive signals from long-distance dependency and negative signals from noises: too small of an $H$ can hardly explore inter-entity relatedness and dependency of long distance, while too large of an $H$ brings much more noises than useful signals, as stated in Section \ref{sec:es}.
        
        	\begin{table}[t]
				\setlength{\abovecaptionskip}{3pt}
            	\centering
            	\caption{The results of $AUC$ w.r.t. different hop numbers.}
                	\begin{tabular}{c|cccc}
                    	\hline
                    	Hop number $H$ & 1 & 2 & 3 & 4 \\
                    	\hline
                    	MovieLens-1M & 0.916 & \textbf{0.919} & 0.915 & 0.918 \\
                    	Book-Crossing & 0.727 & 0.722 & \textbf{0.730} & 0.702 \\
                    	Bing-News & 0.662 & 0.676 & \textbf{0.679} & 0.674 \\
                    	\hline
					\end{tabular}
					\vspace{-0.05in}
				\label{table:hop_number}
			\end{table}

	\subsection{Case Study}
		\begin{figure}[t]
			\centering
  			\includegraphics[width=0.45\textwidth]{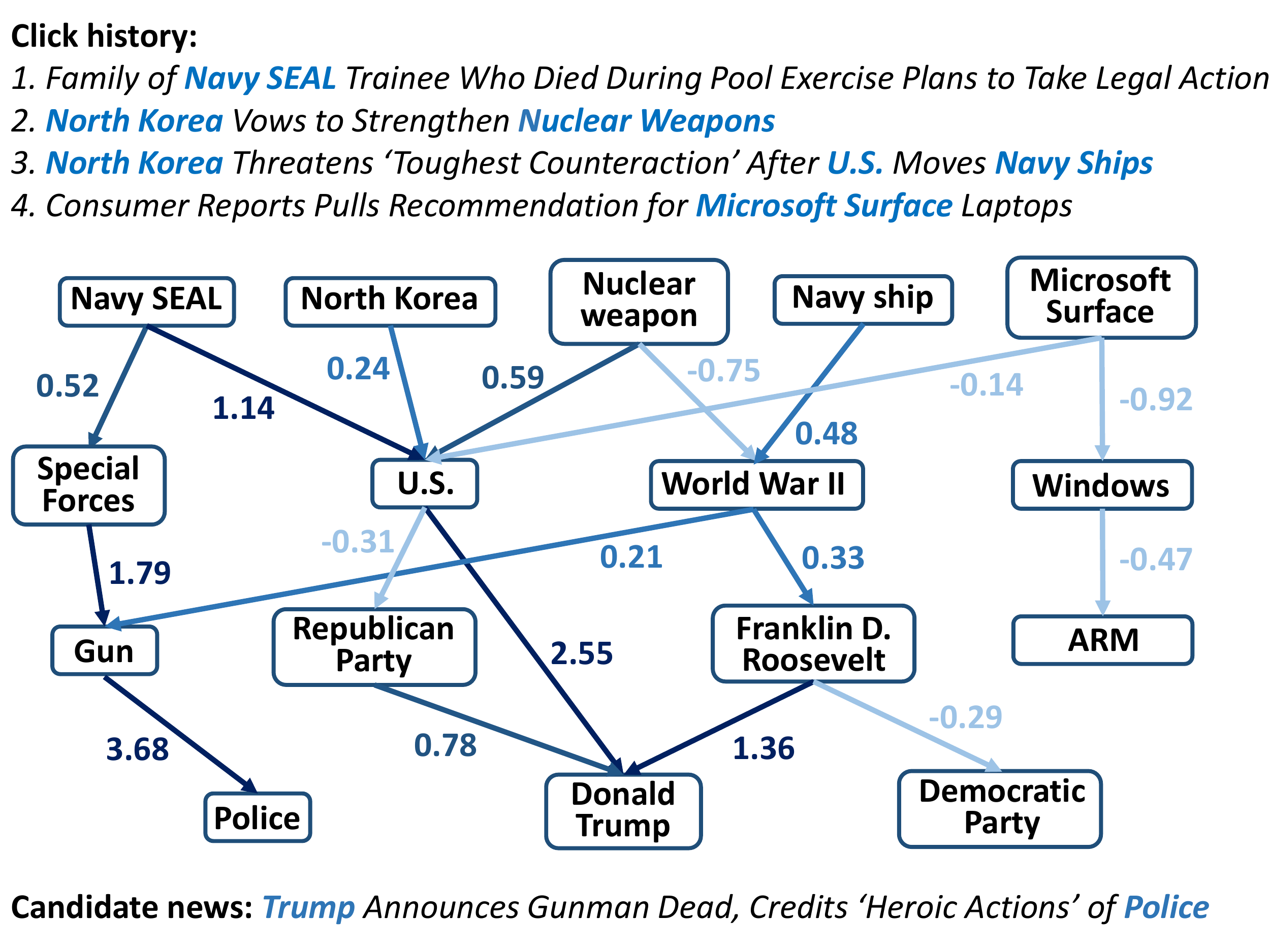}
  			\caption{Visualization of relevance probabilities for a randomly sampled user w.r.t. a piece of candidate news with label $1$. Links with value lower than $-1.0$ are omitted.}
  			\label{fig:cs}
		\end{figure}
		
		To intuitively demonstrate the preference propagation in RippleNet, we randomly sample a user with $4$ clicked pieces of news, and select one candidate news from his test set with label $1$.
		For each of the user's $k$-hop relevant entities, we calculate the (unnormalized) relevance probability between the entity and the candidate news or its $k$-order responses.
		The results are presented in Figure \ref{fig:cs}, in which the darker shade of blue indicates larger values, and we omit names of relations for clearer presentation.
		From Figure \ref{fig:cs} we observe that RippleNet associates the candidate news with the user's relevant entities with different strengths.
		The candidate news can be reached via several paths in the KG with high weights from the user's click history, such as "Navy SEAL"--"Special Forces"--"Gun"--"Police".
		These highlighted paths automatically discovered by preference propagation can thus be used to explain the recommendation result, as discussed in Section \ref{sec:explainability}.
		Additionally, it is also worth noticing that several entities in the KG receive more intensive attention from the user's history, such as "U.S.", "World War II" and "Donald Trump".
		These central entities result from the ripple superposition discussed in Section \ref{sec:ripple_superposition}, and can serve as the user's potential interests for future recommendation.

	\subsection{Parameter Sensitivity}
		\begin{figure}[t]
			\centering
            \begin{subfigure}[b]{0.23\textwidth}
                \includegraphics[width=\textwidth]{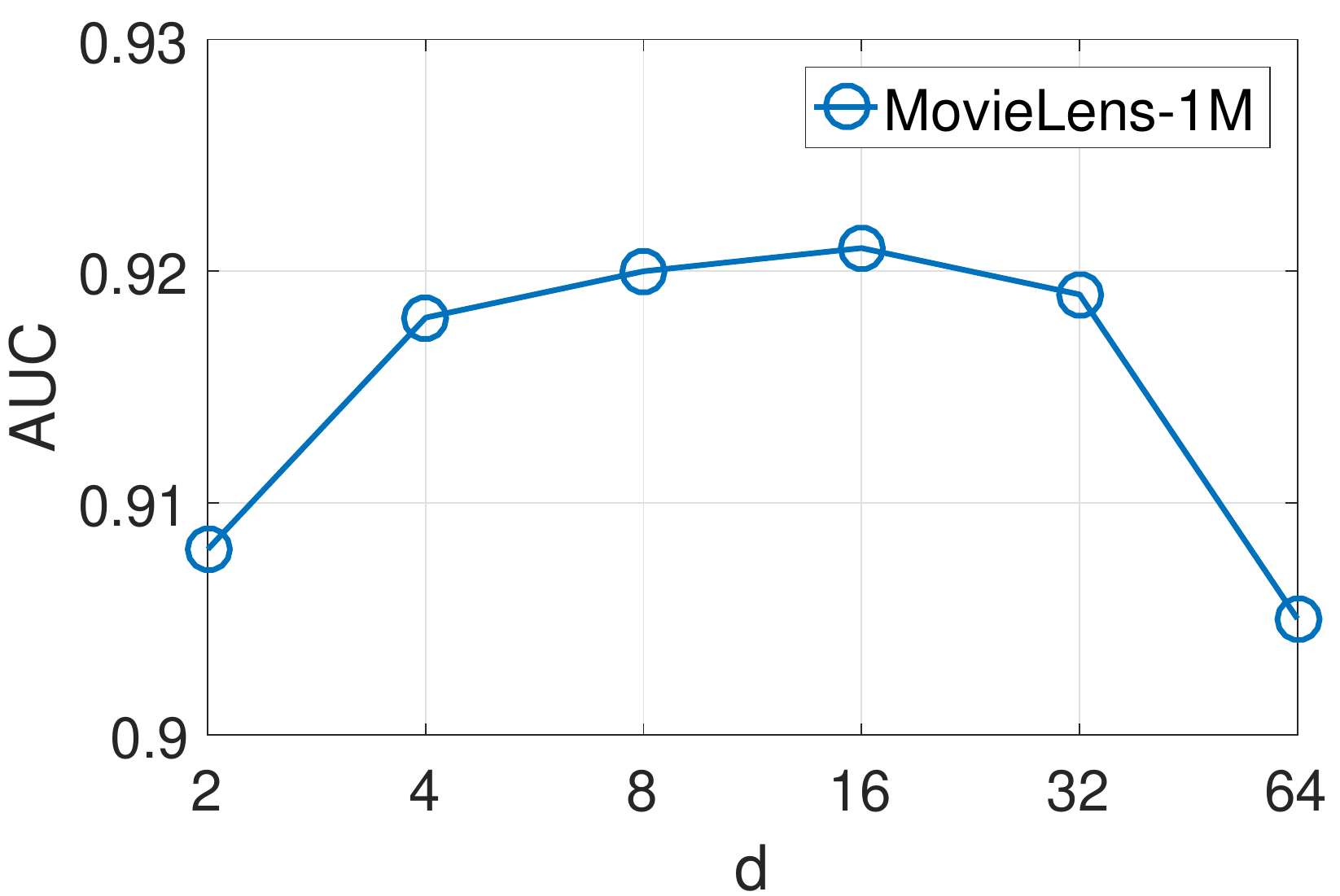}
                \caption{Dimension of embedding}
                \label{fig:ps_1}
            \end{subfigure}
            \hfill
            \begin{subfigure}[b]{0.23\textwidth}
                \includegraphics[width=\textwidth]{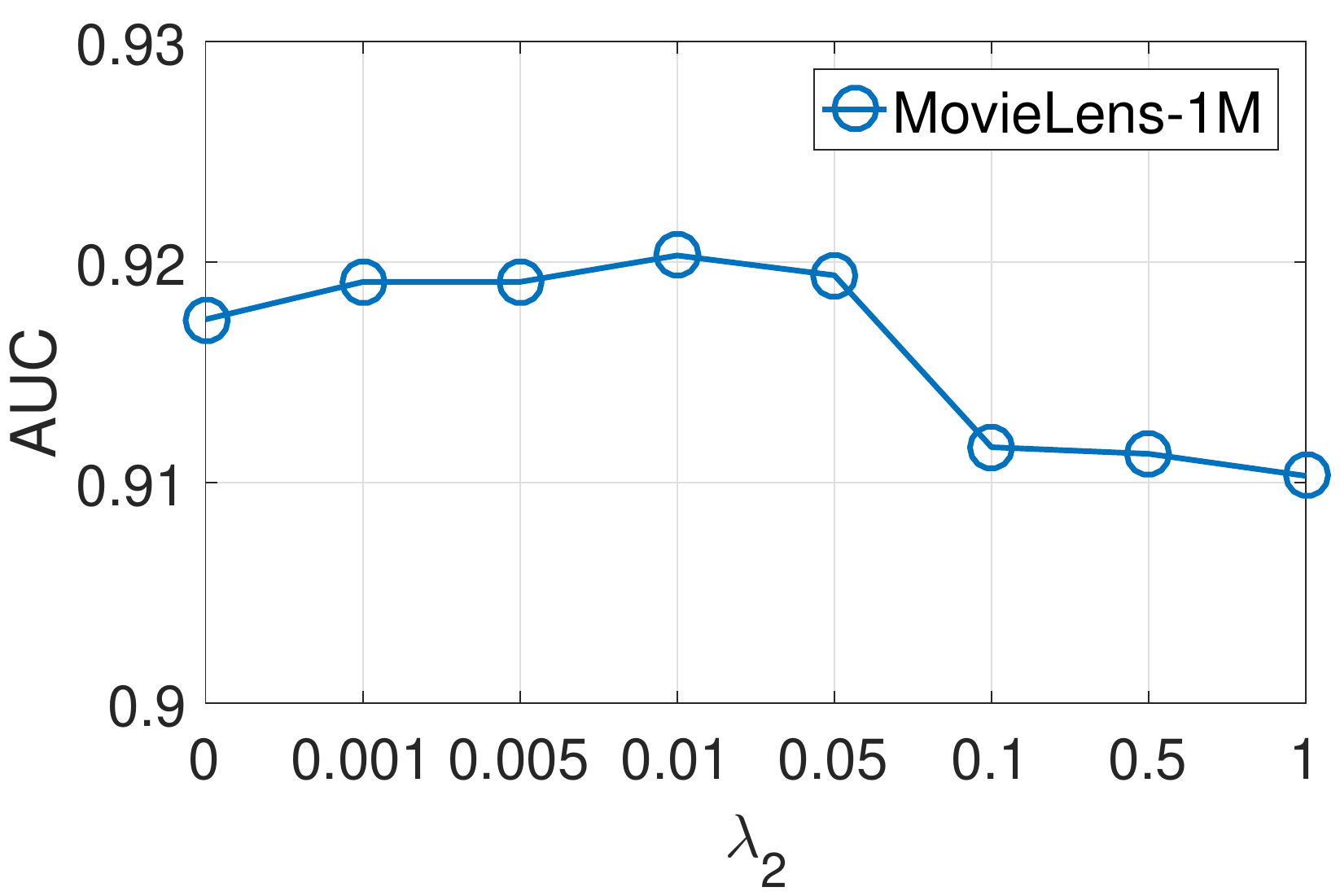}
                \caption{Training weight of KGE term}
                \label{fig:ps_2}
            \end{subfigure}
            \caption{Parameter sensitivity of RippleNet.}
            \label{fig:ps}
        \end{figure}
        
        In this section, we investigate the influence of parameters $d$ and $\lambda_2$ in RippleNet.
        We vary $d$ from $2$ to $64$ and $\lambda_2$ from $0.0$ to $1.0$, respectively, while keeping other parameters fixed.
        The results of $AUC$ on MovieLens-1M are presented in Figure \ref{fig:ps}.
        We observe from Figure \ref{fig:ps_1} that, with the increase of $d$, the performance is boosted at first since embeddings with a larger dimension can encode more useful information, but drops after $d = 16$ due to possible overfitting.
        From Figure \ref{fig:ps_2}, we can see that RippleNet achieves the best performance when $\lambda_2 = 0.01$.
        This is because the KGE term with a small weight cannot provide enough regularization constraints, while a large weight will mislead the objective function.

\section{Conclusion and Future Work}
\label{section_conclusions}
	In this paper, we propose RippleNet, an end-to-end framework that naturally incorporates the knowledge graph into recommender systems.
	RippleNet overcomes the limitations of existing embedding-based and path-based KG-aware recommendation methods by introducing preference propagation, which automatically propagates users' potential preferences and explores their hierarchical interests in the KG.
	RippleNet unifies the preference propagation with regularization of KGE in a Bayesian framework for click-through rate prediction.
	We conduct extensive experiments in three recommendation scenarios.
	The results demonstrate the significant superiority of RippleNet over strong baselines.
	
	For future work, we plan to (1) further investigate the methods of characterizing entity-relation interactions; (2) design non-uniform samplers during preference propagation to better explore users' potential interests and improve the performance.

	
\bibliographystyle{ACM-Reference-Format}
\bibliography{sigproc} 


\begin{thebibliography}{47}


\ifx \showCODEN    \undefined \def \showCODEN     #1{\unskip}     \fi
\ifx \showDOI      \undefined \def \showDOI       #1{#1}\fi
\ifx \showISBNx    \undefined \def \showISBNx     #1{\unskip}     \fi
\ifx \showISBNxiii \undefined \def \showISBNxiii  #1{\unskip}     \fi
\ifx \showISSN     \undefined \def \showISSN      #1{\unskip}     \fi
\ifx \showLCCN     \undefined \def \showLCCN      #1{\unskip}     \fi
\ifx \shownote     \undefined \def \shownote      #1{#1}          \fi
\ifx \showarticletitle \undefined \def \showarticletitle #1{#1}   \fi
\ifx \showURL      \undefined \def \showURL       {\relax}        \fi
\providecommand\bibfield[2]{#2}
\providecommand\bibinfo[2]{#2}
\providecommand\natexlab[1]{#1}
\providecommand\showeprint[2][]{arXiv:#2}

\bibitem[\protect\citeauthoryear{Bahdanau, Cho, and Bengio}{Bahdanau
  et~al\mbox{.}}{2014}]%
        {bahdanau2014neural}
\bibfield{author}{\bibinfo{person}{Dzmitry Bahdanau},
  \bibinfo{person}{Kyunghyun Cho}, {and} \bibinfo{person}{Yoshua Bengio}.}
  \bibinfo{year}{2014}\natexlab{}.
\newblock \showarticletitle{Neural machine translation by jointly learning to
  align and translate}.
\newblock \bibinfo{journal}{\emph{arXiv preprint arXiv:1409.0473}}
  (\bibinfo{year}{2014}).
\newblock


\bibitem[\protect\citeauthoryear{Bauman, Liu, and Tuzhilin}{Bauman
  et~al\mbox{.}}{2017}]%
        {bauman2017aspect}
\bibfield{author}{\bibinfo{person}{Konstantin Bauman}, \bibinfo{person}{Bing
  Liu}, {and} \bibinfo{person}{Alexander Tuzhilin}.}
  \bibinfo{year}{2017}\natexlab{}.
\newblock \showarticletitle{Aspect based recommendations: Recommending items
  with the most valuable aspects based on user reviews}. In
  \bibinfo{booktitle}{\emph{Proceedings of the 23rd ACM SIGKDD International
  Conference on Knowledge Discovery and Data Mining}}. ACM,
  \bibinfo{pages}{717--725}.
\newblock


\bibitem[\protect\citeauthoryear{Bordes, Usunier, Garcia-Duran, Weston, and
  Yakhnenko}{Bordes et~al\mbox{.}}{2013}]%
        {bordes2013translating}
\bibfield{author}{\bibinfo{person}{Antoine Bordes}, \bibinfo{person}{Nicolas
  Usunier}, \bibinfo{person}{Alberto Garcia-Duran}, \bibinfo{person}{Jason
  Weston}, {and} \bibinfo{person}{Oksana Yakhnenko}.}
  \bibinfo{year}{2013}\natexlab{}.
\newblock \showarticletitle{Translating embeddings for modeling
  multi-relational data}. In \bibinfo{booktitle}{\emph{Advances in Neural
  Information Processing Systems}}. \bibinfo{pages}{2787--2795}.
\newblock


\bibitem[\protect\citeauthoryear{Chen, Zhang, He, Nie, Liu, and Chua}{Chen
  et~al\mbox{.}}{2017}]%
        {chen2017attentive}
\bibfield{author}{\bibinfo{person}{Jingyuan Chen}, \bibinfo{person}{Hanwang
  Zhang}, \bibinfo{person}{Xiangnan He}, \bibinfo{person}{Liqiang Nie},
  \bibinfo{person}{Wei Liu}, {and} \bibinfo{person}{Tat-Seng Chua}.}
  \bibinfo{year}{2017}\natexlab{}.
\newblock \showarticletitle{Attentive collaborative filtering: Multimedia
  recommendation with item-and component-level attention}. In
  \bibinfo{booktitle}{\emph{SIGIR}}. ACM, \bibinfo{pages}{335--344}.
\newblock


\bibitem[\protect\citeauthoryear{Chen, Xu, Zhang, Tang, Cao, Qin, and Zha}{Chen
  et~al\mbox{.}}{2018}]%
        {chen2018sequential}
\bibfield{author}{\bibinfo{person}{Xu Chen}, \bibinfo{person}{Hongteng Xu},
  \bibinfo{person}{Yongfeng Zhang}, \bibinfo{person}{Jiaxi Tang},
  \bibinfo{person}{Yixin Cao}, \bibinfo{person}{Zheng Qin}, {and}
  \bibinfo{person}{Hongyuan Zha}.} \bibinfo{year}{2018}\natexlab{}.
\newblock \showarticletitle{Sequential Recommendation with User Memory
  Networks}. In \bibinfo{booktitle}{\emph{Proceedings of the 11th ACM
  International Conference on Web Search and Data Mining}}.
\newblock


\bibitem[\protect\citeauthoryear{Cheng, Koc, Harmsen, Shaked, Chandra, Aradhye,
  Anderson, Corrado, Chai, Ispir, et~al\mbox{.}}{Cheng et~al\mbox{.}}{2016}]%
        {cheng2016wide}
\bibfield{author}{\bibinfo{person}{Heng-Tze Cheng}, \bibinfo{person}{Levent
  Koc}, \bibinfo{person}{Jeremiah Harmsen}, \bibinfo{person}{Tal Shaked},
  \bibinfo{person}{Tushar Chandra}, \bibinfo{person}{Hrishi Aradhye},
  \bibinfo{person}{Glen Anderson}, \bibinfo{person}{Greg Corrado},
  \bibinfo{person}{Wei Chai}, \bibinfo{person}{Mustafa Ispir}, {et~al\mbox{.}}}
  \bibinfo{year}{2016}\natexlab{}.
\newblock \showarticletitle{Wide \& deep learning for recommender systems}. In
  \bibinfo{booktitle}{\emph{Proceedings of the 1st Workshop on Deep Learning
  for Recommender Systems}}. ACM, \bibinfo{pages}{7--10}.
\newblock


\bibitem[\protect\citeauthoryear{Dong, Wei, Zhou, and Xu}{Dong
  et~al\mbox{.}}{2015}]%
        {dong2015question}
\bibfield{author}{\bibinfo{person}{Li Dong}, \bibinfo{person}{Furu Wei},
  \bibinfo{person}{Ming Zhou}, {and} \bibinfo{person}{Ke Xu}.}
  \bibinfo{year}{2015}\natexlab{}.
\newblock \showarticletitle{Question Answering over Freebase with Multi-Column
  Convolutional Neural Networks}. In \bibinfo{booktitle}{\emph{ACL}}.
  \bibinfo{pages}{260--269}.
\newblock


\bibitem[\protect\citeauthoryear{Huang, Zhang, and Huang}{Huang
  et~al\mbox{.}}{2017}]%
        {huang2017mention}
\bibfield{author}{\bibinfo{person}{Haoran Huang}, \bibinfo{person}{Qi Zhang},
  {and} \bibinfo{person}{Xuanjing Huang}.} \bibinfo{year}{2017}\natexlab{}.
\newblock \showarticletitle{Mention Recommendation for Twitter with End-to-end
  Memory Network}. In \bibinfo{booktitle}{\emph{IJCAI}}.
\newblock


\bibitem[\protect\citeauthoryear{Jamali and Ester}{Jamali and Ester}{2010}]%
        {jamali2010matrix}
\bibfield{author}{\bibinfo{person}{Mohsen Jamali} {and} \bibinfo{person}{Martin
  Ester}.} \bibinfo{year}{2010}\natexlab{}.
\newblock \showarticletitle{A matrix factorization technique with trust
  propagation for recommendation in social networks}. In
  \bibinfo{booktitle}{\emph{Proceedings of the 4th ACM conference on
  Recommender systems}}. ACM, \bibinfo{pages}{135--142}.
\newblock


\bibitem[\protect\citeauthoryear{Ji, He, Xu, Liu, and Zhao}{Ji
  et~al\mbox{.}}{2015}]%
        {ji2015knowledge}
\bibfield{author}{\bibinfo{person}{Guoliang Ji}, \bibinfo{person}{Shizhu He},
  \bibinfo{person}{Liheng Xu}, \bibinfo{person}{Kang Liu}, {and}
  \bibinfo{person}{Jun Zhao}.} \bibinfo{year}{2015}\natexlab{}.
\newblock \showarticletitle{Knowledge graph embedding via dynamic mapping
  matrix}. In \bibinfo{booktitle}{\emph{Proceedings of the 53rd Annual Meeting
  of the Association for Computational Linguistics and the 7th International
  Joint Conference on Natural Language Processing (Volume 1: Long Papers)}},
  Vol.~\bibinfo{volume}{1}. \bibinfo{pages}{687--696}.
\newblock


\bibitem[\protect\citeauthoryear{Koren}{Koren}{2008}]%
        {koren2008factorization}
\bibfield{author}{\bibinfo{person}{Yehuda Koren}.}
  \bibinfo{year}{2008}\natexlab{}.
\newblock \showarticletitle{Factorization meets the neighborhood: a
  multifaceted collaborative filtering model}. In
  \bibinfo{booktitle}{\emph{Proceedings of the 14th ACM SIGKDD International
  Conference on Knowledge Discovery and Data Mining}}. ACM,
  \bibinfo{pages}{426--434}.
\newblock


\bibitem[\protect\citeauthoryear{Koren, Bell, and Volinsky}{Koren
  et~al\mbox{.}}{2009}]%
        {koren2009matrix}
\bibfield{author}{\bibinfo{person}{Yehuda Koren}, \bibinfo{person}{Robert
  Bell}, {and} \bibinfo{person}{Chris Volinsky}.}
  \bibinfo{year}{2009}\natexlab{}.
\newblock \showarticletitle{Matrix factorization techniques for recommender
  systems}.
\newblock \bibinfo{journal}{\emph{Computer}} \bibinfo{volume}{42},
  \bibinfo{number}{8} (\bibinfo{year}{2009}).
\newblock


\bibitem[\protect\citeauthoryear{Li, Zhang, Wei, Wu, and Yang}{Li
  et~al\mbox{.}}{2017}]%
        {li2017end}
\bibfield{author}{\bibinfo{person}{Zheng Li}, \bibinfo{person}{Yu Zhang},
  \bibinfo{person}{Ying Wei}, \bibinfo{person}{Yuxiang Wu}, {and}
  \bibinfo{person}{Qiang Yang}.} \bibinfo{year}{2017}\natexlab{}.
\newblock \showarticletitle{End-to-End Adversarial Memory Network for
  Cross-domain Sentiment Classification}. In
  \bibinfo{booktitle}{\emph{Proceedings of the 26th International Joint
  Conference on Artificial Intelligence}}.
\newblock


\bibitem[\protect\citeauthoryear{Lin, Liu, Sun, Liu, and Zhu}{Lin
  et~al\mbox{.}}{2015}]%
        {lin2015learning}
\bibfield{author}{\bibinfo{person}{Yankai Lin}, \bibinfo{person}{Zhiyuan Liu},
  \bibinfo{person}{Maosong Sun}, \bibinfo{person}{Yang Liu}, {and}
  \bibinfo{person}{Xuan Zhu}.} \bibinfo{year}{2015}\natexlab{}.
\newblock \showarticletitle{Learning Entity and Relation Embeddings for
  Knowledge Graph Completion}. In \bibinfo{booktitle}{\emph{AAAI}}.
  \bibinfo{pages}{2181--2187}.
\newblock


\bibitem[\protect\citeauthoryear{Liu, Wu, and Yang}{Liu et~al\mbox{.}}{2017}]%
        {liu2017analogical}
\bibfield{author}{\bibinfo{person}{Hanxiao Liu}, \bibinfo{person}{Yuexin Wu},
  {and} \bibinfo{person}{Yiming Yang}.} \bibinfo{year}{2017}\natexlab{}.
\newblock \showarticletitle{Analogical Inference for Multi-Relational
  Embeddings}. In \bibinfo{booktitle}{\emph{Proceedings of the 34th
  International Conference on Machine Learning}}. \bibinfo{pages}{2168--2178}.
\newblock


\bibitem[\protect\citeauthoryear{Mikolov, Sutskever, Chen, Corrado, and
  Dean}{Mikolov et~al\mbox{.}}{2013}]%
        {mikolov2013distributed}
\bibfield{author}{\bibinfo{person}{Tomas Mikolov}, \bibinfo{person}{Ilya
  Sutskever}, \bibinfo{person}{Kai Chen}, \bibinfo{person}{Greg~S Corrado},
  {and} \bibinfo{person}{Jeff Dean}.} \bibinfo{year}{2013}\natexlab{}.
\newblock \showarticletitle{Distributed representations of words and phrases
  and their compositionality}. In \bibinfo{booktitle}{\emph{Advances in Neural
  Information Processing Systems}}. \bibinfo{pages}{3111--3119}.
\newblock


\bibitem[\protect\citeauthoryear{Miller, Fisch, Dodge, Karimi, Bordes, and
  Weston}{Miller et~al\mbox{.}}{2016}]%
        {miller2016key}
\bibfield{author}{\bibinfo{person}{Alexander Miller}, \bibinfo{person}{Adam
  Fisch}, \bibinfo{person}{Jesse Dodge}, \bibinfo{person}{Amir-Hossein Karimi},
  \bibinfo{person}{Antoine Bordes}, {and} \bibinfo{person}{Jason Weston}.}
  \bibinfo{year}{2016}\natexlab{}.
\newblock \showarticletitle{Key-value memory networks for directly reading
  documents}.
\newblock \bibinfo{journal}{\emph{arXiv preprint arXiv:1606.03126}}
  (\bibinfo{year}{2016}).
\newblock


\bibitem[\protect\citeauthoryear{Mnih, Heess, Graves, et~al\mbox{.}}{Mnih
  et~al\mbox{.}}{2014}]%
        {mnih2014recurrent}
\bibfield{author}{\bibinfo{person}{Volodymyr Mnih}, \bibinfo{person}{Nicolas
  Heess}, \bibinfo{person}{Alex Graves}, {et~al\mbox{.}}}
  \bibinfo{year}{2014}\natexlab{}.
\newblock \showarticletitle{Recurrent models of visual attention}. In
  \bibinfo{booktitle}{\emph{Advances in Neural Information Processing
  Systems}}. \bibinfo{pages}{2204--2212}.
\newblock


\bibitem[\protect\citeauthoryear{Nickel, Rosasco, Poggio, et~al\mbox{.}}{Nickel
  et~al\mbox{.}}{2016}]%
        {nickel2016holographic}
\bibfield{author}{\bibinfo{person}{Maximilian Nickel}, \bibinfo{person}{Lorenzo
  Rosasco}, \bibinfo{person}{Tomaso~A Poggio}, {et~al\mbox{.}}}
  \bibinfo{year}{2016}\natexlab{}.
\newblock \showarticletitle{Holographic Embeddings of Knowledge Graphs}. In
  \bibinfo{booktitle}{\emph{AAAI}}. \bibinfo{pages}{1955--1961}.
\newblock


\bibitem[\protect\citeauthoryear{Rendle}{Rendle}{2012}]%
        {rendle2012factorization}
\bibfield{author}{\bibinfo{person}{Steffen Rendle}.}
  \bibinfo{year}{2012}\natexlab{}.
\newblock \showarticletitle{Factorization machines with libfm}.
\newblock \bibinfo{journal}{\emph{ACM Transactions on Intelligent Systems and
  Technology (TIST)}} \bibinfo{volume}{3}, \bibinfo{number}{3}
  (\bibinfo{year}{2012}), \bibinfo{pages}{57}.
\newblock


\bibitem[\protect\citeauthoryear{Rockt{\"a}schel, Singh, and
  Riedel}{Rockt{\"a}schel et~al\mbox{.}}{2015}]%
        {rocktaschel2015injecting}
\bibfield{author}{\bibinfo{person}{Tim Rockt{\"a}schel},
  \bibinfo{person}{Sameer Singh}, {and} \bibinfo{person}{Sebastian Riedel}.}
  \bibinfo{year}{2015}\natexlab{}.
\newblock \showarticletitle{Injecting logical background knowledge into
  embeddings for relation extraction}. In \bibinfo{booktitle}{\emph{Proceedings
  of the 2015 Conference of the North American Chapter of the Association for
  Computational Linguistics: Human Language Technologies}}.
  \bibinfo{pages}{1119--1129}.
\newblock


\bibitem[\protect\citeauthoryear{Seo, Huang, Yang, and Liu}{Seo
  et~al\mbox{.}}{2017}]%
        {seo2017interpretable}
\bibfield{author}{\bibinfo{person}{Sungyong Seo}, \bibinfo{person}{Jing Huang},
  \bibinfo{person}{Hao Yang}, {and} \bibinfo{person}{Yan Liu}.}
  \bibinfo{year}{2017}\natexlab{}.
\newblock \showarticletitle{Interpretable convolutional neural networks with
  dual local and global attention for review rating prediction}. In
  \bibinfo{booktitle}{\emph{Proceedings of the Eleventh ACM Conference on
  Recommender Systems}}. ACM, \bibinfo{pages}{297--305}.
\newblock


\bibitem[\protect\citeauthoryear{Sharma and Cosley}{Sharma and Cosley}{2013}]%
        {sharma2013social}
\bibfield{author}{\bibinfo{person}{Amit Sharma} {and} \bibinfo{person}{Dan
  Cosley}.} \bibinfo{year}{2013}\natexlab{}.
\newblock \showarticletitle{Do social explanations work?: studying and modeling
  the effects of social explanations in recommender systems}. In
  \bibinfo{booktitle}{\emph{Proceedings of the 22nd international conference on
  World Wide Web}}. ACM, \bibinfo{pages}{1133--1144}.
\newblock


\bibitem[\protect\citeauthoryear{Sukhbaatar, Weston, Fergus,
  et~al\mbox{.}}{Sukhbaatar et~al\mbox{.}}{2015}]%
        {sukhbaatar2015end}
\bibfield{author}{\bibinfo{person}{Sainbayar Sukhbaatar},
  \bibinfo{person}{Jason Weston}, \bibinfo{person}{Rob Fergus},
  {et~al\mbox{.}}} \bibinfo{year}{2015}\natexlab{}.
\newblock \showarticletitle{End-to-end memory networks}. In
  \bibinfo{booktitle}{\emph{Advances in Neural Information Processing
  Systems}}. \bibinfo{pages}{2440--2448}.
\newblock


\bibitem[\protect\citeauthoryear{Sun, Yuan, Xie, McDonald, and Zhang}{Sun
  et~al\mbox{.}}{2017}]%
        {sun2017collaborative}
\bibfield{author}{\bibinfo{person}{Yu Sun}, \bibinfo{person}{Nicholas~Jing
  Yuan}, \bibinfo{person}{Xing Xie}, \bibinfo{person}{Kieran McDonald}, {and}
  \bibinfo{person}{Rui Zhang}.} \bibinfo{year}{2017}\natexlab{}.
\newblock \showarticletitle{Collaborative Intent Prediction with Real-Time
  Contextual Data}.
\newblock \bibinfo{journal}{\emph{ACM Transactions on Information Systems}}
  \bibinfo{volume}{35}, \bibinfo{number}{4} (\bibinfo{year}{2017}),
  \bibinfo{pages}{30}.
\newblock


\bibitem[\protect\citeauthoryear{Tai, Socher, and Manning}{Tai
  et~al\mbox{.}}{2015}]%
        {tai2015improved}
\bibfield{author}{\bibinfo{person}{Kai~Sheng Tai}, \bibinfo{person}{Richard
  Socher}, {and} \bibinfo{person}{Christopher~D Manning}.}
  \bibinfo{year}{2015}\natexlab{}.
\newblock \showarticletitle{Improved Semantic Representations From
  Tree-Structured Long Short-Term Memory Networks}. In
  \bibinfo{booktitle}{\emph{Proceedings of the 53rd Annual Meeting of the
  Association for Computational Linguistics and the 7th International Joint
  Conference on Natural Language Processing (Volume 1: Long Papers)}},
  Vol.~\bibinfo{volume}{1}. \bibinfo{pages}{1556--1566}.
\newblock


\bibitem[\protect\citeauthoryear{Tintarev and Masthoff}{Tintarev and
  Masthoff}{2007}]%
        {tintarev2007survey}
\bibfield{author}{\bibinfo{person}{Nava Tintarev} {and} \bibinfo{person}{Judith
  Masthoff}.} \bibinfo{year}{2007}\natexlab{}.
\newblock \showarticletitle{A survey of explanations in recommender systems}.
  In \bibinfo{booktitle}{\emph{IEEE 23rd International Conference on Data
  Engineering Workshop}}. IEEE, \bibinfo{pages}{801--810}.
\newblock


\bibitem[\protect\citeauthoryear{Trouillon, Welbl, Riedel, Gaussier, and
  Bouchard}{Trouillon et~al\mbox{.}}{2016}]%
        {trouillon2016complex}
\bibfield{author}{\bibinfo{person}{Th{\'e}o Trouillon},
  \bibinfo{person}{Johannes Welbl}, \bibinfo{person}{Sebastian Riedel},
  \bibinfo{person}{{\'E}ric Gaussier}, {and} \bibinfo{person}{Guillaume
  Bouchard}.} \bibinfo{year}{2016}\natexlab{}.
\newblock \showarticletitle{Complex embeddings for simple link prediction}. In
  \bibinfo{booktitle}{\emph{International Conference on Machine Learning}}.
  \bibinfo{pages}{2071--2080}.
\newblock


\bibitem[\protect\citeauthoryear{Vig, Sen, and Riedl}{Vig
  et~al\mbox{.}}{2009}]%
        {vig2009tagsplanations}
\bibfield{author}{\bibinfo{person}{Jesse Vig}, \bibinfo{person}{Shilad Sen},
  {and} \bibinfo{person}{John Riedl}.} \bibinfo{year}{2009}\natexlab{}.
\newblock \showarticletitle{Tagsplanations: explaining recommendations using
  tags}. In \bibinfo{booktitle}{\emph{Proceedings of the 14th international
  conference on Intelligent user interfaces}}. ACM, \bibinfo{pages}{47--56}.
\newblock


\bibitem[\protect\citeauthoryear{Wang, Wang, Wang, Zhao, Zhang, Zhang, Xie, and
  Guo}{Wang et~al\mbox{.}}{2018a}]%
        {wang2018graphgan}
\bibfield{author}{\bibinfo{person}{Hongwei Wang}, \bibinfo{person}{Jia Wang},
  \bibinfo{person}{Jialin Wang}, \bibinfo{person}{Miao Zhao},
  \bibinfo{person}{Weinan Zhang}, \bibinfo{person}{Fuzheng Zhang},
  \bibinfo{person}{Xing Xie}, {and} \bibinfo{person}{Minyi Guo}.}
  \bibinfo{year}{2018}\natexlab{a}.
\newblock \showarticletitle{Graphgan: Graph representation learning with
  generative adversarial nets}. In \bibinfo{booktitle}{\emph{AAAI}}.
  \bibinfo{pages}{2508--2515}.
\newblock


\bibitem[\protect\citeauthoryear{Wang, Wang, Zhao, Cao, and Guo}{Wang
  et~al\mbox{.}}{2017c}]%
        {wang2017joint}
\bibfield{author}{\bibinfo{person}{Hongwei Wang}, \bibinfo{person}{Jia Wang},
  \bibinfo{person}{Miao Zhao}, \bibinfo{person}{Jiannong Cao}, {and}
  \bibinfo{person}{Minyi Guo}.} \bibinfo{year}{2017}\natexlab{c}.
\newblock \showarticletitle{Joint Topic-Semantic-aware Social Recommendation
  for Online Voting}. In \bibinfo{booktitle}{\emph{Proceedings of the 2017 ACM
  Conference on Information and Knowledge Management}}. ACM,
  \bibinfo{pages}{347--356}.
\newblock


\bibitem[\protect\citeauthoryear{Wang, Zhang, Hou, Xie, Guo, and Liu}{Wang
  et~al\mbox{.}}{2018b}]%
        {wang2018shine}
\bibfield{author}{\bibinfo{person}{Hongwei Wang}, \bibinfo{person}{Fuzheng
  Zhang}, \bibinfo{person}{Min Hou}, \bibinfo{person}{Xing Xie},
  \bibinfo{person}{Minyi Guo}, {and} \bibinfo{person}{Qi Liu}.}
  \bibinfo{year}{2018}\natexlab{b}.
\newblock \showarticletitle{Shine: Signed heterogeneous information network
  embedding for sentiment link prediction}. In
  \bibinfo{booktitle}{\emph{Proceedings of the Eleventh ACM International
  Conference on Web Search and Data Mining}}. ACM, \bibinfo{pages}{592--600}.
\newblock


\bibitem[\protect\citeauthoryear{Wang, Zhang, Xie, and Guo}{Wang
  et~al\mbox{.}}{2018c}]%
        {wang2018dkn}
\bibfield{author}{\bibinfo{person}{Hongwei Wang}, \bibinfo{person}{Fuzheng
  Zhang}, \bibinfo{person}{Xing Xie}, {and} \bibinfo{person}{Minyi Guo}.}
  \bibinfo{year}{2018}\natexlab{c}.
\newblock \showarticletitle{DKN: Deep Knowledge-Aware Network for News
  Recommendation}. In \bibinfo{booktitle}{\emph{Proceedings of the 2018 World
  Wide Web Conference on World Wide Web}}. International World Wide Web
  Conferences Steering Committee, \bibinfo{pages}{1835--1844}.
\newblock


\bibitem[\protect\citeauthoryear{Wang, Wang, Zhang, and Yan}{Wang
  et~al\mbox{.}}{2017b}]%
        {wang2017combining}
\bibfield{author}{\bibinfo{person}{Jin Wang}, \bibinfo{person}{Zhongyuan Wang},
  \bibinfo{person}{Dawei Zhang}, {and} \bibinfo{person}{Jun Yan}.}
  \bibinfo{year}{2017}\natexlab{b}.
\newblock \showarticletitle{Combining Knowledge with Deep Convolutional Neural
  Networks for Short Text Classification}. In
  \bibinfo{booktitle}{\emph{IJCAI}}.
\newblock


\bibitem[\protect\citeauthoryear{Wang, Mao, Wang, and Guo}{Wang
  et~al\mbox{.}}{2017a}]%
        {wang2017knowledge}
\bibfield{author}{\bibinfo{person}{Quan Wang}, \bibinfo{person}{Zhendong Mao},
  \bibinfo{person}{Bin Wang}, {and} \bibinfo{person}{Li Guo}.}
  \bibinfo{year}{2017}\natexlab{a}.
\newblock \showarticletitle{Knowledge graph embedding: A survey of approaches
  and applications}.
\newblock \bibinfo{journal}{\emph{IEEE Transactions on Knowledge and Data
  Engineering}} \bibinfo{volume}{29}, \bibinfo{number}{12}
  (\bibinfo{year}{2017}), \bibinfo{pages}{2724--2743}.
\newblock


\bibitem[\protect\citeauthoryear{Wang, Yu, Ren, Tao, Zhang, Yu, and Wang}{Wang
  et~al\mbox{.}}{2017d}]%
        {wang2017dynamic}
\bibfield{author}{\bibinfo{person}{Xuejian Wang}, \bibinfo{person}{Lantao Yu},
  \bibinfo{person}{Kan Ren}, \bibinfo{person}{Guanyu Tao},
  \bibinfo{person}{Weinan Zhang}, \bibinfo{person}{Yong Yu}, {and}
  \bibinfo{person}{Jun Wang}.} \bibinfo{year}{2017}\natexlab{d}.
\newblock \showarticletitle{Dynamic attention deep model for article
  recommendation by learning human editors' demonstration}. In
  \bibinfo{booktitle}{\emph{Proceedings of the 23rd ACM SIGKDD International
  Conference on Knowledge Discovery and Data Mining}}. ACM,
  \bibinfo{pages}{2051--2059}.
\newblock


\bibitem[\protect\citeauthoryear{Wang, Zhang, Feng, and Chen}{Wang
  et~al\mbox{.}}{2014}]%
        {wang2014knowledge}
\bibfield{author}{\bibinfo{person}{Zhen Wang}, \bibinfo{person}{Jianwen Zhang},
  \bibinfo{person}{Jianlin Feng}, {and} \bibinfo{person}{Zheng Chen}.}
  \bibinfo{year}{2014}\natexlab{}.
\newblock \showarticletitle{Knowledge Graph Embedding by Translating on
  Hyperplanes}. In \bibinfo{booktitle}{\emph{AAAI}}.
  \bibinfo{pages}{1112--1119}.
\newblock


\bibitem[\protect\citeauthoryear{Weston, Chopra, and Bordes}{Weston
  et~al\mbox{.}}{2014}]%
        {weston2014memory}
\bibfield{author}{\bibinfo{person}{Jason Weston}, \bibinfo{person}{Sumit
  Chopra}, {and} \bibinfo{person}{Antoine Bordes}.}
  \bibinfo{year}{2014}\natexlab{}.
\newblock \showarticletitle{Memory networks}.
\newblock \bibinfo{journal}{\emph{arXiv preprint arXiv:1410.3916}}
  (\bibinfo{year}{2014}).
\newblock


\bibitem[\protect\citeauthoryear{Xie, Liu, and Sun}{Xie et~al\mbox{.}}{2016}]%
        {xie2016representation}
\bibfield{author}{\bibinfo{person}{Ruobing Xie}, \bibinfo{person}{Zhiyuan Liu},
  {and} \bibinfo{person}{Maosong Sun}.} \bibinfo{year}{2016}\natexlab{}.
\newblock \showarticletitle{Representation Learning of Knowledge Graphs with
  Hierarchical Types.}. In \bibinfo{booktitle}{\emph{IJCAI}}.
  \bibinfo{pages}{2965--2971}.
\newblock


\bibitem[\protect\citeauthoryear{Xu, Bai, Bian, Gao, Wang, Liu, and Liu}{Xu
  et~al\mbox{.}}{2014}]%
        {xu2014rc}
\bibfield{author}{\bibinfo{person}{Chang Xu}, \bibinfo{person}{Yalong Bai},
  \bibinfo{person}{Jiang Bian}, \bibinfo{person}{Bin Gao},
  \bibinfo{person}{Gang Wang}, \bibinfo{person}{Xiaoguang Liu}, {and}
  \bibinfo{person}{Tie-Yan Liu}.} \bibinfo{year}{2014}\natexlab{}.
\newblock \showarticletitle{Rc-net: A general framework for incorporating
  knowledge into word representations}. In
  \bibinfo{booktitle}{\emph{Proceedings of the 23rd ACM International
  Conference on Conference on Information and Knowledge Management}}. ACM,
  \bibinfo{pages}{1219--1228}.
\newblock


\bibitem[\protect\citeauthoryear{Yang, Yih, He, Gao, and Deng}{Yang
  et~al\mbox{.}}{2015}]%
        {yang2015embedding}
\bibfield{author}{\bibinfo{person}{Bishan Yang}, \bibinfo{person}{Wen-tau Yih},
  \bibinfo{person}{Xiaodong He}, \bibinfo{person}{Jianfeng Gao}, {and}
  \bibinfo{person}{Li Deng}.} \bibinfo{year}{2015}\natexlab{}.
\newblock \showarticletitle{Embedding entities and relations for learning and
  inference in knowledge bases}. In \bibinfo{booktitle}{\emph{Proceedings of
  the 3rd International Conference on Learning Representations}}.
\newblock


\bibitem[\protect\citeauthoryear{Yu, Ren, Sun, Gu, Sturt, Khandelwal, Norick,
  and Han}{Yu et~al\mbox{.}}{2014}]%
        {yu2014personalized}
\bibfield{author}{\bibinfo{person}{Xiao Yu}, \bibinfo{person}{Xiang Ren},
  \bibinfo{person}{Yizhou Sun}, \bibinfo{person}{Quanquan Gu},
  \bibinfo{person}{Bradley Sturt}, \bibinfo{person}{Urvashi Khandelwal},
  \bibinfo{person}{Brandon Norick}, {and} \bibinfo{person}{Jiawei Han}.}
  \bibinfo{year}{2014}\natexlab{}.
\newblock \showarticletitle{Personalized entity recommendation: A heterogeneous
  information network approach}. In \bibinfo{booktitle}{\emph{Proceedings of
  the 7th ACM International Conference on Web Search and Data Mining}}.
  \bibinfo{pages}{283--292}.
\newblock


\bibitem[\protect\citeauthoryear{Zhang, Yuan, Lian, Xie, and Ma}{Zhang
  et~al\mbox{.}}{2016}]%
        {zhang2016collaborative}
\bibfield{author}{\bibinfo{person}{Fuzheng Zhang},
  \bibinfo{person}{Nicholas~Jing Yuan}, \bibinfo{person}{Defu Lian},
  \bibinfo{person}{Xing Xie}, {and} \bibinfo{person}{Wei-Ying Ma}.}
  \bibinfo{year}{2016}\natexlab{}.
\newblock \showarticletitle{Collaborative knowledge base embedding for
  recommender systems}. In \bibinfo{booktitle}{\emph{Proceedings of the 22nd
  ACM SIGKDD International Conference on Knowledge Discovery and Data Mining}}.
  ACM, \bibinfo{pages}{353--362}.
\newblock


\bibitem[\protect\citeauthoryear{Zhang, Lai, Zhang, Zhang, Liu, and Ma}{Zhang
  et~al\mbox{.}}{2014}]%
        {zhang2014explicit}
\bibfield{author}{\bibinfo{person}{Yongfeng Zhang}, \bibinfo{person}{Guokun
  Lai}, \bibinfo{person}{Min Zhang}, \bibinfo{person}{Yi Zhang},
  \bibinfo{person}{Yiqun Liu}, {and} \bibinfo{person}{Shaoping Ma}.}
  \bibinfo{year}{2014}\natexlab{}.
\newblock \showarticletitle{Explicit factor models for explainable
  recommendation based on phrase-level sentiment analysis}. In
  \bibinfo{booktitle}{\emph{Proceedings of the 37th international ACM SIGIR
  conference on Research \& development in information retrieval}}. ACM,
  \bibinfo{pages}{83--92}.
\newblock


\bibitem[\protect\citeauthoryear{Zhao, Yao, Li, Song, and Lee}{Zhao
  et~al\mbox{.}}{2017}]%
        {zhao2017meta}
\bibfield{author}{\bibinfo{person}{Huan Zhao}, \bibinfo{person}{Quanming Yao},
  \bibinfo{person}{Jianda Li}, \bibinfo{person}{Yangqiu Song}, {and}
  \bibinfo{person}{Dik~Lun Lee}.} \bibinfo{year}{2017}\natexlab{}.
\newblock \showarticletitle{Meta-graph based recommendation fusion over
  heterogeneous information networks}. In \bibinfo{booktitle}{\emph{Proceedings
  of the 23rd ACM SIGKDD International Conference on Knowledge Discovery and
  Data Mining}}. ACM, \bibinfo{pages}{635--644}.
\newblock


\bibitem[\protect\citeauthoryear{Zhong, Zhang, Wang, Wan, and Chen}{Zhong
  et~al\mbox{.}}{2015}]%
        {zhong2015aligning}
\bibfield{author}{\bibinfo{person}{Huaping Zhong}, \bibinfo{person}{Jianwen
  Zhang}, \bibinfo{person}{Zhen Wang}, \bibinfo{person}{Hai Wan}, {and}
  \bibinfo{person}{Zheng Chen}.} \bibinfo{year}{2015}\natexlab{}.
\newblock \showarticletitle{Aligning knowledge and text embeddings by entity
  descriptions}. In \bibinfo{booktitle}{\emph{Proceedings of the 2015
  Conference on Empirical Methods in Natural Language Processing}}.
  \bibinfo{pages}{267--272}.
\newblock


\bibitem[\protect\citeauthoryear{Zhou, Song, Zhu, Ma, Yan, Dai, Zhu, Jin, Li,
  and Gai}{Zhou et~al\mbox{.}}{2017}]%
        {zhou2017deep}
\bibfield{author}{\bibinfo{person}{Guorui Zhou}, \bibinfo{person}{Chengru
  Song}, \bibinfo{person}{Xiaoqiang Zhu}, \bibinfo{person}{Xiao Ma},
  \bibinfo{person}{Yanghui Yan}, \bibinfo{person}{Xingya Dai},
  \bibinfo{person}{Han Zhu}, \bibinfo{person}{Junqi Jin}, \bibinfo{person}{Han
  Li}, {and} \bibinfo{person}{Kun Gai}.} \bibinfo{year}{2017}\natexlab{}.
\newblock \showarticletitle{Deep interest network for click-through rate
  prediction}.
\newblock \bibinfo{journal}{\emph{arXiv preprint arXiv:1706.06978}}
  (\bibinfo{year}{2017}).
\newblock


\end{thebibliography}

\end{document}